%% file: paper.tex
\documentclass[11pt]{article}
\pdfoutput=1
\usepackage{verbatim}
\usepackage{amsmath}
\usepackage{amsfonts}
\usepackage{amssymb}
\usepackage{bbm}
\usepackage{latexsym}
\usepackage{lscape} 
\usepackage{epsfig}
\usepackage{color}
\usepackage{graphicx}
\usepackage{subfigure}

\usepackage{hyperref}

\usepackage{multirow}
\usepackage{mathrsfs}

\usepackage{amscd}
\usepackage{cite}

\usepackage[small,sc]{caption}

\input{texdefs}

\begin{document}

\thispagestyle{empty}

\begin{flushright}
LMU-ASC 60/13 
\\
\end{flushright}
\vskip 2 cm
\begin{center}
{\Large {\bf MSSM--like models on $\boldsymbol{\Intr_8}$ toroidal orbifolds}
}
\\[0pt]

\bigskip
\bigskip {\large
{\bf Stefan Groot Nibbelink$^{a,}$}\footnote{
E-mail: Groot.Nibbelink@physik.uni-muenchen.de},
{\bf Orestis Loukas$^{a,b,}$}\footnote{
E-mail: O.Loukas@physik.uni-muenchen.de},
\bigskip }\\[0pt]
\vspace{0.23cm}
${}^a$ {\it 
Arnold Sommerfeld Center for Theoretical Physics,\\
~~Ludwig-Maximilians-Universit\"at M\"unchen, 80333 M\"unchen, Germany
}
\\[1ex]
${}^b$ {\it 
Physics Department, National Technical University of Athens,\\
Zografou Campus, 15780 Athens, Greece
}
\\[1ex] 
\bigskip
\end{center}

\input{Abstract}

\newpage 
\setcounter{page}{1}
\tableofcontents

\input{intro}

\input{geom}
\input{strings}

\input{Z8geoms}

\input{models}

\bibliographystyle{paper}
{\small
\bibliography{literatur}
}

\end{document}

%% file: texdefs.tex
\definecolor{gray}{gray}{.5}

\DeclareMathSymbol{\mg}{\mathrel}{symbols}{"1D}

\newcommand{\ga}{\alpha}
\newcommand{\gb}{\beta}

\newcommand{\gm}{\mu}

\newcommand{\gth}{\theta}

\newcommand{\gs}{\sigma}
\newcommand{\gt}{\tau}

\newcommand{\gp}{\pi}

\newcommand{\gG}{\Gamma}

\newcommand{\gL}{\Lambda}

\newcommand{\gPs}{\Psi}

\newcommand{\ra}{\rightarrow}

\newcommand{\beq}{\begin{equation}}
\newcommand{\eeq}{\end{equation}}
\newcommand{\barr}{\begin{array}}
\newcommand{\earr}{\end{array}}
\newcommand{\equ}[1]{\begin{gather} #1 \end{gather}}
\newcommand{\equa}[1]{\begin{align} #1 \end{align}}

\newcommand{\enums}[1]{\begin{enumerate} #1 \end{enumerate}}
\newcommand{\tabu}[2]{\begin{tabular}{#1} #2 \end{tabular}}
\newcommand{\arry}[2]{\begin{array}{#1} #2 \end{array}}

\newcommand{\sfrac}[2]{\mbox{$\frac{#1}{#2}$}}
\newcounter{oldcounter}

\newcommand{\Intr}{\mathbb{Z}}
\newcommand{\Cplx}{\mathbb{C}}
\newcommand{\Real}{\mathbb{R}}

\newcommand{\ba}[2]{\[\begin{array}{#2}\label{#1}}
\newcommand{\ea}{\end{array}\]}
\newcommand{\be}{\begin{equation}}
\newcommand{\ee}{\end{equation}}
\newcommand{\bea}{\begin{eqnarray}}
\newcommand{\eea}{\end{eqnarray}}



%% file: Abstract.tex
\subsection*{\centering Abstract}

We extend the promising heterotic string searches for MSSM--like models to $\Intr_8$ orbifolds. 
There exist five inequivalent $\Intr_8$ toroidal orbifolds distinguished by two types of twists that act on five different torus lattices; 
one of which cannot be represented as a Lie--algebra root lattice. 
Contrary to previous investigations, we study the consequences of the different underlying orbifold torus lattices.
Therefore, rather than focussing on one particular geometry, we perform systematic model searches on these five $\Intr_8$ orbifolds simultaneously,  taking all possible inequivalent $SU(5)$ and $SO(10)$ gauge shifts as our starting point. 
We present cumulative Tables and Figures comparing the chiral SM and vector--like exotic spectra on these geometries.


%% file: intro.tex
 \section{Introduction}
 \label{sc:intro}

%
The aim of string phenomenology is to build string models which can be related to the real world, i.e.\ the Standard Model (SM) of Particle Physics. In order to have computational control one often makes various further assumptions on both the string background and the resulting effective field theory. In particular, supersymmetry has proven very helpful for string model building, hence, many groups have focussed on the search for the (Minimal) Supersymmetric extension of the Standard Model (MSSM). The heterotic string~\cite{Gross:1984dd} has been in particular quite successful in producing MSSM models following essentially three different approaches:  

%
The first approach is  heterotic model building on smooth (often elliptically fibered) Calabi--Yau spaces with non--Abelian vector bundles~\cite{Candelas:1985en}. This has resulted in MSSM--like models~\cite{Bouchard:2005ag} with possible supersymmetry breaking built in~\cite{Braun:2005ux,Braun:2005bw,Braun:2005nv}. These constructions were hampered by the fact that it is very challenging to construct stable vector bundles. More recently, this problem has been by--passed by using line bundles instead~\cite{Anderson:2011ns,Anderson:2012yf}. However, since  these approaches correspond to complicated interacting worldsheet theories, studies of such Calabi--Yau compactifications typically start from the supergravity approximation of the full heterotic string. 

%
The other approaches to heterotic model building take exact Conformal Field Theory (CFT) descriptions as their starting points. The first MSSM--like heterotic string models were obtained within the so--called free--fermionic formulation~\cite{Faraggi:1989ka}. In addition, a large set of MSSM models were constructed as interacting CFTs, like the Gepner models~\cite{Dijkstra:2004ym} and the rational conformal field theory models~\cite{Dijkstra:2004cc}.

%
Also heterotic orbifolds~\cite{Dixon:1985jw,Dixon:1986jc,Ibanez:1986tp,Ibanez:1987pj} can be used to construct MSSM--like models. For a comprehensive introduction to these models see e.g.~\cite{string_compactification_phys_rept,Choi2006}. In Refs.~\cite{Buchmuller:2005jr,Buchmuller:2006ik,Lebedev:2006kn,Lebedev:2008un} more than two hundred MSSM--like models have been assembled on the orbifold $T^6/{\mathbb Z}_\text{6--II}$. There have been similar so--called mini--landscape searches performed on the $T^6/\Intr_2\times \Intr_4$ orbifold~\cite{Z2xZ4}. In addition, there have been models constructed on the $T^6/\Intr_\text{12--I}$ orbifold~\cite{Kim:2006hv,Kim:2007mt}. Most of these models break the $E_8\times E_8$ gauge group of the heterotic string directly down to the SM gauge group via appropriate gauge shifts and discrete Wilson lines. This means that when these models are being fully resolved, the hyper--charge will be broken~\cite{Blowup}. This can be avoided by using a non--local Wilson line to perform the breaking of the $SU(5)$ Grand Unified (GUT) gauge group to the SM~\cite{Z2xZ2_Nibbelink}.

\subsection*{Purpose of present paper}

%
This paper extends the search for MSSM--like heterotic orbifold models in a new direction: Contrary to previous studies we would like to investigate the consequences of the different underlying torus lattices. To this end, we analyze heterotic models constructed on $T^6/\Intr_8$ orbifolds. (In the past there has been some GUT model work~\cite{Kobayashi2,Kawabe:1994mj} done on some of the $T^6/\Intr_\text{8}$ orbifolds.) There are two types of $\Intr_8$ actions often referred to as $\Intr_\text{8--I}$ and $\Intr_\text{8--II}$, see e.g.\ Ref.~\cite{shiftZ8}. Recently there has been a complete classification of all possible supersymmetric toroidal orbifolds~\cite{nonlocal_patrick}; a previous investigation~\cite{string_compactification_phys_rept} aimed to classify all Abelian orbifolds on Lie--lattices. For $\Intr_8$ orbifolds these results mean that there are in total five orbifolds: The $\Intr_\text{8--I}$ action admits three lattices, two of which are Lie--algebra lattices while the third is not,\footnote{
Ref.~\cite{string_compactification_phys_rept} identified two Lie--lattice, for $\Intr_\text{8--I}$:   $SO(9)\times SO(5)$ and $SO(8)^{[2]}\times SO(5)$; Ref.~\cite{nonlocal_patrick} finds that both are equivalent and, furthermore, that there exists two additional lattices. We will show that one of these lattices can be interpreted as the $SU(4)\times SU(4)$ Lie--lattice. We suspect that this lattice was missed in the classification of Ref.~\cite{string_compactification_phys_rept} since the orbifold action involves an outer--automorphism which interchanges both $SU(4)$ factors.}
while the $\Intr_\text{8--II}$ has two inequivalent Lie--lattices. 
 
%
In this paper we systematically investigate heterotic $\Intr_8$ orbifolds on these five inequivalent torus lattices. In particular, we give detailed expositions of their fixed point and tori structures. We present computer--aided MSSM--like model searches based on configurations with gauge shifts that break the observable $E_8$ down to the $SU(5)$ GUT group directly. In addition, we consider models in which the gauge shifts first break the observable gauge group to $SO(10)$ while discrete Wilson lines subsequently reduce the gauge group to $SU(5)$. By scanning over all possible remaining compatible Wilson lines, we find in total $753$ models on the five $\Intr_8$ orbifolds. For these models we present average values for the number of vector--like Higgses and exotics that these models possess. In addition, we display how both SM matter as well as exotics are distributed over the various twisted sectors of the five $\Intr_8$ orbifolds in histograms.

\subsection*{Paper outline}

In order to make our paper self--contained we begin with a brief review of $T^6/\Intr_N$ orbifold geometries in Section~\ref{sc:geom} using a language that applies equally well to orbifolds based on factorizable and non--factorizable tori with underlying Lie-- or non--Lie--lattices. In Section~\ref{sc:strings} we briefly consider the necessary ingredients to describe heterotic strings on such orbifolds. In particular, we introduce the gauge shift and discrete Wilson lines as the input data for heterotic orbifold models. 

From Section~\ref{sc:Z8geoms} onwards we focus specifically on heterotic model building on $\Intr_8$ orbifolds. Section~\ref{sc:Z8geoms} describes the three $\Intr_\text{8--I}$ and the two $\Intr_\text{8--II}$ orbifold geometries and indicates the consequences for the possible Wilson lines. We begin Section~\ref{sc:models} outlining the way we set up our model search on the $\Intr_8$ orbifolds. In the next Subsections we summarize the results of our model scans and present Tables and Figures comparing the particle content on the five inequivalent $\Intr_8$ orbifolds.

\subsection*{Acknowledgements}

We would like to thank Patrick Vaudrevange for many very helpful discussions and communications. We would also like to thank Saul Ramos--Sanchez for extensive communications about model searches on the $\Intr_\text{8--I}$ orbifold on the torus with the $SO(9)\times SO(5)$ lattice. On behalf of the LMU physics library we would like to thank K.S.\ Choi for donating his book. 
We should also express our gratitude to the DAAD for the programme ``Vollstipendium f\"{u}r Absolventen von deutschen Auslandsschulen'' within the ``PASCH--Initiative''. This work has been supported by the LMUExcellent Programme.

%% file: geom.tex
\section{Orbifold geometries}
\label{sc:geom}

In this section we set the notation for our description of $\Intr_N$ orbifolds.

\subsection{six--tori}
\label{sc:NonFact}

%
A six dimensional torus $T^6 = \Real^6/\Gamma$ is defined by specifying a six dimensional lattice
\begin{equation}
\Gamma = \lbrace \hspace{0.03cm} n_\alpha\, e_\alpha, \text{ } n_\alpha \in \mathbb{Z}~, 
\quad \alpha = 1,...,6\rbrace~. 
\end{equation}
This lattice defines how the points in $\mathbb{R}^6$ are identified on the six--torus when they differ by some lattice vectors, i.e.\ 
\begin{equation}
x \sim x+n_\alpha\, e_\alpha~, 
\end{equation}
for some integers $n_\ga \in \Intr$. The basis vectors $e_\alpha$ can often be conveniently chosen to be the simple roots of some semi--simple Lie--algebra of rank six. 

%
The Cartesian inner product of these basis vectors defines the torus or Gram metric,  
\begin{equation} \label{torus metric}
G_{\alpha \beta} = e_\alpha \cdot e^{\;}_\beta~, 
\end{equation} 
where $\cdot$ denotes the standard Euclidean inner product. If this metric can be brought to a form consisting of three $2\times 2$ blocks on the diagonal by a change of basis, we call the lattice factorizable; otherwise we say it is non--factorizable. When the orbifold is based on a Lie--lattice we also refer to the Gram metric as the Cartan matrix.

\subsection[${\Intr_N}$ orbifolds]{$\mathbf{\Intr_N}$ orbifolds}

%
On the torus lattice $\gG$ we act with an orbifold twist $\theta: \gG \ra \gG$ that generates the point group $P$. In this paper we restrict ourselves to Abelian point groups that are isomorphic to $\Intr_8$, but we leave $\Intr_N$ general for now. The orbifold action can be diagonalized over the complex space $\Cplx^3$, with complex coordinates $z = (z_1,z_2,z_3)$, so that we can represent the orbifold twist as 
\begin{equation} \label{form of theta}
\theta = \text{diag}(\text{e}^{2 \pi i \upsilon^1},  \text{e}^{2 \pi i \upsilon^2},  \text{e}^{2 \pi i \upsilon^3})~. 
\end{equation} 
The entries of the twist vector
\(
\upsilon = (0, \upsilon^1, \upsilon^2, \upsilon^3)
\)
specify the rotation angles of the discrete rotation in the three complex planes and are therefore quantized in units of $1/N$ for a $\Intr_N$ orbifold. (The first zero entry has been introduced for later convenience when describing the heterotic string on such spaces.) In order that the orbifold preserves $\mathcal{N}=1$ supersymmetry in four dimensions we impose the Calabi--Yau condition:
\begin{equation}  \label{Calaby_Yau_condition}
\upsilon^1 + \upsilon^2 + \upsilon^3 = 0~. 
\end{equation}
The orbifold is defined as points on the six--torus identified up to point group transformations: 
\begin{equation}
\mathbb{O} = T^6/P = \mathbb{R}^6 / S~.
\end{equation}
Here $S$ denotes the space group that combines lattice translations with orbifold twists. A general element $g=(\gth^k, n_\ga e_\ga)$ of $S$ acts as 
\begin{equation}
g\circ x=\theta^k\, x + n_\alpha\, e_\alpha~. 
\end{equation}
%

%
In order to avoid confusion during calculations, we denote the action of the twist $\gth$ on the real lattice basis vectors, $e_\ga$, by the matrix $Q$ defined as: 
\equ{\label{LatticeComp} 
\gth\, e_\ga = e_\gb \, Q^\gb{}_\ga~. 
}
The matrix $Q$ needs to be an element of $GL(6, \Intr)$, i.e.\ $Q$ has to be invertible and both $Q$ and $Q^{-1}$ are required to have only integral entries. The metric $G$ is compatible with the orbifold symmetry generated by $\gth$, provided that 
\begin{equation}\label{scalar product invariance for g} 
Q^T \, G \, Q = G~. 
\end{equation}

%% file: strings.tex
\section{Heterotic string on $\mathbf{\Intr_N}$ orbifolds}
\label{sc:strings}

\subsection{Worldsheet boundary conditions} 
\label{sc:BCs} 

%
The string worldsheet is conventionally parameterized by a space $\gs$ and time $\gt$ coordinates. The closed strings of the heterotic string are characterized by the periodicity conditions in the $\gs$--direction. For the coordinate fields $X^\gm$ these read
\equ{ 
X (\tau, \sigma + \pi ) = g\circ X (\tau, \sigma) 
= \gth^k\, X(\tau,\sigma) + n_\ga\, e_\ga~.  
}
%
%
The right--moving superpartners $\gPs_R^\mu$ of the coordinate fields $X^\mu$ can can be pair--wise bosonized to right--moving bosons $H_R^i$, where $i=0,1,2,3$ in light--cone--gauge. Their boundary conditions read 
\equ{ 
H_R(\gt, \gs+\gp) = g\circ H_R(\gt,\gs) 
= H_R(\gt,\gs) + \pi\, q_\text{sh}~, 
\qquad 
q_\text{sh} = q + v_g~, 
}
introducing the local twist $v_g = k\, v$ and $q$ denotes either vectorial or spinorial weights of the light--cone Lorentz group $SO(8)$, 
\begin{equation} \label{q_8_V_S}
q = \big(\, \underline{\pm 1, 0, 0, 0} \,\big)
\quad \text{or} \quad 
q = \big( \pm\sfrac{1}{2}, \pm\sfrac{1}{2}, \pm\sfrac{1}{2}, \pm\sfrac{1}{2} \big)~, 
\end{equation}
with an even number of minus signs, respectively. With the underlined entries we denote all possible permutations of these entries. 
%
%
In addition the heterotic string contains sixteen left--moving coordinates $\smash{X_L^I}$, $I=1,\ldots, 16$, living on a torus $T^{16}$ of radius $\smash{R = 1/\sqrt{2}}$ (in string units) defining the 
closed string gauge degrees of freedom. The boundary condition for these left--moving coordinates are given by
\begin{equation} \label{shifted_momentum} 
X_L (\tau , \sigma + \pi ) = g \circ X_L(\tau, \sigma) 
= X_L (\tau , \sigma ) + \pi\, p_\text{sh}~, 
\qquad 
p_\text{sh} = p+V_g~, 
\end{equation}
in terms of the so--called shifted momentum $p_\text{sh}$. In the definition of the shifted momentum we introduced the concept of the local gauge shift vector, 
\equ{ \label{gauge shift} 
V_g = k\, V + n_\alpha\, W_\alpha~,
}
referring to the gauge shift at a particular fixed point on the orbifold. (At the origin one recovers the standard shift: $V_{(\gth,0)} = V$ since $k=1$ and $n_\ga=0$ then.) It is defined in terms of the (gauge) shift, $V$, and discrete Wilson lines, $W_\alpha$, associated to the orbifold element $\gth$ and torus--translations $e_\ga$, respectively.

%
Finally, in \eqref{shifted_momentum} the shifted momentum, $p_\text{sh}$, is given in terms of the lattice weight $p \in \gL$. The lattice $\gL$ corresponds to the root lattice of $E_8\times E_8$. A single $E_8$ root lattice may be characterized as 
\begin{equation}
E_8 = \Big\lbrace 
\big(n_1, n_2, ... , n_8\big) \oplus 
\big(n_1+\sfrac{1}{2}, n_2+\sfrac{1}{2}, ..., n_8+\sfrac{1}{2}\big) 
~\Big\vert~ n_i \in \mathbb{Z}, \sum n_i \in 2 \mathbb{Z} 
\Big\rbrace~.
\end{equation}
This lattice is generated by the $E_8$ roots of norm $2$ given by 
\begin{equation} \label{root lattice E8}
\big(\,\underline{\pm 1, \pm 1, 0^6} \,\big) 
\quad \text{and} \quad 
\big( \pm \sfrac{1}{2}^8\, \big)~ 
\text{with even number of minus signs}~.
\end{equation}
%

%
Finally, $N$ is the order of the gauge shift $V$ of discrete orbifold group $\Intr_N$: 
\equ{
N\, V \cong 0~. 
}
Here $a \cong b$ means that the difference of the vectors $a, b$ is a lattice vector of $\gL$. 
The orders $N_\ga$ of the discrete Wilson lines $W_\ga$ depend on the matrix $Q$ and hence ultimately on the torus lattice $\gG$ and are determined by the requirements, 
\equ{
W_\ga \cong 
W_\gb\, Q^\gb{}_\ga~. 
}
This is just a reflection of the compatibility condition \eqref{LatticeComp} of the torus lattice with the orbifold twist. These conditions do not only determine the orders of the discrete Wilson lines but often also require that various Wilson lines are related to each other.

\subsection{Modular invariance conditions}
\label{sc:ModInv}

%
A crucial consistency condition of heterotic orbifolded strings is the requirement that the one--loop partition function be modular. In addition, one has to ensure that the model involves well--defined GSO and orbifold projections~\cite{vaccumphase}. Together this gives a set of stringent consistency conditions, which e.g.\ safeguards the resulting effective target space theory from dangerous anomalies. 

%
The resulting conditions on gauge shift $V$ and discrete Wilson lines $W_\alpha$ for a $\mathbb{Z}_N$ orbifold can be represented as 
\begin{subequations} 
\label{ModInv} 
\equa{ 
N\, (V^2 - \upsilon^2) = 
N_\alpha\, W_\alpha \cdot W_\alpha &
= 0 \ \text{mod}\ 2 ~, 
\label{W_a x W_a}
\\[2ex] 
N_\alpha\, W_\alpha \cdot V = 
\text{gcd}( N_\alpha, N_\beta ) \, W_\alpha \cdot W_\beta & 
= 0 \ \text{mod}\ 2~, ~~ \alpha \neq \beta~, 
\label{V x W_a}
}
\end{subequations} 
here no summation over $\alpha, \beta$ implied.

\subsection{Massless string excitations} 
\label{sc:MasslessStates}

The left-- and right--moving masses of the string are given by 
\begin{equation} \label{LRmasses}
\frac{M_L^2}{8} = 
\frac{p_\text{sh}^2}{2} 
+ \tilde{N} - 1 + \delta \tilde{c}_g
\quad \text{and} \quad 
\frac{M_R}{8} = 
\frac{q_\text{sh}^2}{2} 
+N - \frac{1}{2} + \delta \tilde{c}_g~,
\end{equation}
respectively. Here, the shift in the zero point energy, $\delta \tilde{c}_g$, is defined as follows: In terms of  $\tilde\upsilon_g = \upsilon_g~\text{mod}~1$ such that $0 \leq \tilde\upsilon_g^i < 1$, the shift in the zero point energy is given by
\begin{equation} \label{ZeroPointE}
\delta \tilde{c}_g =   \frac{1}{2} \sum_i 
\tilde\upsilon_g^i  (1 - \tilde\upsilon_g^i  )~. 
\end{equation}
The massless string excitations are selected by requiring that the left-- and right--moving masses vanish
\begin{equation}\label{massless}
M_L^2 = M_R^2 =0~. 
\end{equation}
The first equality is the so--called level matching condition.

Not all massless states defined by these equations are part of the physical spectrum, only those that survive the orbifold projection are. Let $h$ be a space group element that commutes with $g$, i.e.\ $[h,g]=0$. Since the action of $h$ should leave the surviving massless states invariant, they should acquire trivial phases under $h$. This is guaranteed when: 
\begin{equation} \label{Projection condition for commuting g, h}
p_{sh}\cdot V_h - R \cdot \upsilon_h =
\frac 12\, \big( V_g\cdot V_h - \upsilon_g \cdot \upsilon_h \big) 
~\text{mod} ~1~, 
\end{equation}
where 
$R^i = q_{sh}^i - \tilde{N}^i + \tilde{N}^{\overline{i}}$.

The projection conditions~\eqref{Projection condition for commuting g, h} are in particular important to determine the unbroken gauge group in four dimensions. As the gauge fields come from the untwisted sector, these conditions reduce to 
\equ{ \label{unbrokenGaugeGroup} 
p \cdot V=0~\text{mod}~1 
\quad \text{and} \quad 
 p \cdot W_\alpha=0~\text{mod}~1~, 
}
for the $E_8\times E_8$ roots $p\in \gL$.

%% file: Z8geoms.tex
%
\newcommand{\ZeightIFixedSets}{
\begin{table}[ht!!] 
\caption{This Table lists the locations of the fixed points and two--tori on the $T^6/\Intr_\text{8--I}$ orbifolds for the three inequivalent lattices. The first, second and third twisted sectors live on fixed points; the fourth twisted sector consists of fixed two--tori. The lattice vectors that span these fixed two--tori are also indicated. 
\label{FixedPointsZ8I}}
\begin{center}
\scalebox{.92}{
\renewcommand{\arraystretch}{1.2}
\begin{tabular}{|c||c|c|c|}
\hline
\textbf{Twist} & \multicolumn{3}{|c|}{\textbf{$\mathbf{\Intr_\text{8--I}}$ fixed sets on $\mathbf{T^6}$ with}}  
\\
& \textbf{$\mathbf{SO(9)\times SO(5)}$ lattice} & \textbf{$\mathbf{SU(4)\times SU(4)}$ lattice} & 
\textbf{non--Lie lattice}
\\\hline\hline
\multirow{3}{*}{$\theta\,,\, \theta^3$} & 
 $\frac 12(e_1 + e_3 + e_5)\,,$
 &$\frac 34 e_1 + \frac 12 e_2 + \frac 14 e_3 + \frac 34 e_4 + \frac 12 e_5 + \frac 14 e_6\,,$ 
& $\frac 12(e_1+e_2+e_3+e_4)\,,$ \\
&$\frac 12(e_1+e_3)\,,\, \frac 12 e_5\,,\, \underline{0}\,,$
& $\frac 14 e_1 + \frac 12 e_2 + \frac 34 e_3 + \frac 14 e_4 + \frac 12 e_5 + \frac 34 e_6\,,$ 
& $\frac 12(e_2+e_4+e_5+e_6)\,,$  \\
& 
& $\frac 12(e_1+e_3+e_4+e_6)\,,\, \underline{0}$ 
& $\frac 12(e_1+e_3+e_5+e_6)\,,\, \underline{0}$ \\
\hline
\multirow{14}{*}{$\theta^2$} 
& $\frac 12(e_1 + e_3 + e_5+e_6)\,,$
&$\frac 34 e_1 + \frac 12 e_2 + \frac 14 e_3 + \frac 34 e_4 + \frac 12 e_5 + \frac 14 e_6\,,$ 
& $\frac 12(e_1+e_2+e_3+e_4+e_5+e_6)\,,$ 
\\
&  $\frac 12(e_2 + e_3 + e_5 + e_6)\,,$
&$\frac 34 e_1 + \frac 12 e_2 + \frac 14 e_3 + \frac 14 e_4 + \frac 12 e_5 + \frac 34 e_6\,,$
& $\frac 14(e_1+e_2+3e_3+3e_4) + \frac 12 e_5\,,$
\\
& $ \frac 12(e_1 + e_2 + e_5 + e_6)\,,$
& $\frac 14 e_1 + \frac 12 e_2 + \frac 34 e_3 + \frac 34 e_4 + \frac 12 e_5 + \frac 14 e_6\,,$ 
& $\frac 14(e_1+e_2+3e_3+3e_4) + \frac 12 e_6\,,$ 
\\
& $\frac 12(e_1 + e_3 + e_5)\,,$
& $\frac 14 e_1 + \frac 12 e_2 + \frac 34 e_3 + \frac 14 e_4 + \frac 12 e_5 + \frac 34 e_6\,,$  
& $\frac 14(e_1+3e_2+3e_3+e_4) + \frac 12 e_5\,,$ 
\\
& $\frac 12(e_2 + e_3 + e_5)\,,$
&  $\frac 12 (e_1 + e_3 + e_5) + \frac 14 e_4 + \frac 34 e_6\,,$
& $\frac 14(e_1+3e_2+3e_3+e_4) + \frac 12 e_6\,,$
\\
&  $ \frac 12(e_1 + e_2 + e_5)\,,$ 
& $\frac 12 (e_1 + e_3 + e_5) + \frac 34 e_4 + \frac 14 e_6\,,$ 
& $\frac 14(3e_1+e_2+e_3+3e_4) + \frac 12 e_5\,,$  
\\
&  $\frac 12(e_1 + e_3 +e_6)\,,$
& $\frac 14 e_1 + \frac 12 (e_2+e_4+e_6) + \frac 34 e_3\,,$
& $\frac 14(3e_1+e_2+e_3+3e_4) + \frac 12 e_6\,,$ 
\\
& $\frac 12(e_2 + e_3 + e_6)\,,$
& $\frac 34 e_1 + \frac 12 (e_2+e_4+e_6) + \frac 14 e_3\,,$ 
& $\frac 14(3e_1+3e_2+e_3+e_4) + \frac 12 e_5\,,$  
\\
& $ \frac 12(e_1 + e_2 + e_6)\,,$
&  $\frac 12 (e_1+e_3+e_4+e_6)\,,$
&  $\frac 14(3e_1+3e_2+e_3+e_4) + \frac 12 e_6\,,$ 
\\ 
& $\frac 12(e_1 + e_3)\,,\, \frac 12(e_2 + e_3)\,,$
& $\frac 14 e_1 + \frac 12 e_2 + \frac 34 e_3\,,$ 
& $\frac 12(e_1+e_2+e_3+e_4)\,,$ 
\\
& $ \frac 12(e_1 + e_2)\,,\, \frac 12(e_5+e_6)\,,$ 
&  $\frac 14 e_4 + \frac 12 e_5 + \frac 34 e_6\,,$  
&  $\frac 12(e_1+e_3+e_5+e_6)\,,$ 
\\
& $\frac 12 e_5\,,\, \frac 12 e_6\,,\, \underline{0}$
& $\frac 34 e_1 + \frac 12 e_2 + \frac 14 e_3\,,$
& $\frac 12(e_2+e_4+e_5+e_6)\,,$ 
\\
& 
& $\frac 34 e_4 + \frac 12 e_5 + \frac 14 e_6\,,$
&  $\frac 12(e_1+e_3)\,,\, \frac 12(e_2+e_4)\,,$
\\ 
& 
&  $\frac 12 (e_1+e_3)\,,\, \frac 12 (e_4+e_6)\,,\, \underline{0}$
&  $\frac 12(e_5+e_6)\,,\, \underline{0}$
\\\hline \hline 
\multirow{13}{*}{$\theta^4$} 
& \multicolumn{2}{|c}{\textbf{Fixed two--tori spanned by the lattice vectors}}  & $\mathbf{e_5-e_6\,,}$ 
\\
& $\mathbf{e_5\,,\, e_6}$
& $\mathbf{e_1+e_3\,,\,e_4+e_6}$ 
& $\mathbf{e_1+e_2+e_3+e_4-e_5-e_6}$ 
\\
\cline{2-4}\cline{2-4} 
& $\frac 12(e_1+e_2+e_3+e_4)\,,$ 
&  $\frac 12(e_1 + e_2 + e_4 + e_5)\,,$
& $\frac 12(e_1+e_2+e_3+e_4)\,,$ 
\\
&$\frac 12(e_1+e_3+e_4)\,,$ 
& $\frac 12 (e_2 + e_3 + e_5 + e_6)\,,$ 
& $\frac 12(e_2+e_4+e_5+e_6)\,,$ 
\\
& $\frac 12(e_1+e_2+e_4)\,,$ 
& $\frac 12 (e_2 + e_3 + e_4 + e_5)\,,$ 
& $\frac 12(e_1+e_3+e_5+e_6)\,,$ 
\\
& $\frac 12(e_1+e_2+e_3)\,,$ 
& $\frac 12 (e_1 + e_2 + e_5 + e_6)\,,$ 
& $\frac 12(e_1+e_2+e_4)\,,$ 
\\
& $\frac 12(e_2+e_3+e_4)\,,$
& $\frac 12 (e_1+e_2)\,,\,\frac 12 (e_2+e_3)\,,$
& $\frac 12(e_1+e_3+e_4)\,,$ 
\\
& $\frac 12(e_1+e_2)\,,\,\frac 12(e_2+e_3)\,,$ 
& $\frac 12 (e_4+e_5)\,,\,\frac 12 (e_5+e_6)\,,\,\underline{0}$
& $\frac 12(e_1+e_2+e_3)\,,$
\\
& $\frac 12(e_2+e_4)\,,\,\frac 12(e_1+e_4)\,,$ 
&
& $\frac 12(e_2+e_3+e_4)\,,$ 
\\
& $\frac 12(e_1+e_3)\,,\,\frac 12(e_3+e_4)\,,$
&
& $\frac 12(e_1+e_2)\,,\,\frac 12(e_2+e_3)\,,$
\\
& $\frac 12e_1\,,\, \frac 12e_2\,,\, \frac 12e_3\,,\, \frac 12e_4 \,,\,\underline{0}$
& 
& $\frac 12(e_3+e_4)\,,\,\frac 12(e_1+e_4)\,,$ 
\\
& 
& 
& $\frac 12(e_1+e_3)\,,\,\frac 12(e_2+e_4)\,,$ 
\\
&  
& 
& $\frac 12e_1\,,\, \frac 12e_2\,,\,\frac 12e_3\,,\,\frac 12e_4\,,\,\underline{0}$ 
\\ \hline 
\end{tabular}
}
\end{center}
\end{table}
}
%

%
\newcommand{\ZeightIIFixedSets}{ 
\begin{table}[ht] 
\caption{This Table lists the locations of the fixed points and two--tori on the $T^6/\Intr_\text{8--II}$ orbifolds for the two inequivalent lattices. The first and third twisted sectors live on fixed points; the second and fourth twisted sector consist of fixed two--tori. The lattice vectors that span these fixed two--tori are indicated as well. 
\label{FixedPointsZ8II}}
\begin{center}
\scalebox{1}{
\renewcommand{\arraystretch}{1.2}
\begin{tabular}{|c||c|c|}
\hline
\textbf{Twist} & \multicolumn{2}{|c|}{\textbf{$\mathbf{\Intr_\text{8--II}}$ fixed sets on $\mathbf{T^6}$ with}}  
\\
& \textbf{$\mathbf{SO(9)\times SU(2)^2}$ lattice} & \textbf{$\mathbf{SO(10)\times SU(2)}$ lattice} 
\\\hline\hline
\multirow{5}{*}{$\theta\,,\, \theta^3$} 
&  $\frac 12(e_1 + e_3 + e_5+e_6)\,,$
& $\frac 12(e_1 + e_3) \pm \frac 14(e_4 - e_5) + \frac 12 e_6\,,$   \\ 
&  $\frac 12(e_1 + e_3 + e_5)\,,$
& $\frac 12(e_1 + e_3) \pm \frac 14(e_4 - e_5)\,,$ \\ 
&  $\frac 12(e_1 + e_3+e_6)\,,$
& $\frac 12(e_4 + e_5 + e_6)\,,$\\ 
&  $\frac 12(e_1 + e_3)\,,\, \frac 12(e_5+e_6)$
& $\frac 12(e_4 + e_5)\,,\, \frac 12 e_6\,,\, \underline{0}$\\ 
&  $\frac 12 e_5\,,\, \frac 12 e_6\,,\, \underline{0}$
& $$ \\ 
\hline \hline 
\multirow{4}{*}{$\theta^2$} 
& \multicolumn{2}{|c|}{\textbf{Fixed two--tori spanned by the lattice vectors}} 
\\
& $\mathbf{e_5\,,\, e_6}$
& $\mathbf{e_4-e_5\,,\,e_6}$ 
\\
\cline{2-3}\cline{2-3}
& $\frac 12(e_1 + e_3)\,,\,  \frac 12(e_2 + e_3)\,,$ 
& $\frac 12(e_1 + e_3 + e_4 + e_5)\,,$ \\
& $\frac 12(e_1 + e_2)\,,\, \underline{0}$
& $\frac 12(e_1 + e_3)\,,\, \frac 12(e_4 + e_5)\,,\, \underline{0}$ 
\\ \hline  \hline 
\multirow{13}{*}{$\theta^4$} 
& \multicolumn{2}{|c|}{\textbf{Fixed two--tori spanned by the lattice vectors}} 
\\
& $\mathbf{e_5\,,\, e_6}$
& $\mathbf{e_4-e_5\,,\,e_6}$ 
\\
\cline{2-3}\cline{2-3}
& $ \frac 12(e_1 + e_2 + e_3 + e_4)\,,$ 
& $\frac 12(e_1 + e_2 + e_3 + e_4 + e_5)\,,$ \\
& $ \frac 12(e_1 + e_2 + e_3)\,,$ 
& $\frac 12(e_1 + e_3 + e_4 + e_5)\,,$  \\
& $ \frac 12(e_2 + e_3 + e_4)\,,$ 
& $\frac 12(e_1 + e_2 + e_4 + e_5)\,,$\\
& $ \frac 12(e_1 + e_2 + e_4)\,,$ 
& $\frac 12 (e_2 + e_3 + e_4 + e_5)\,,$ \\
& $ \frac 12(e_1 + e_3 + e_4)\,,$ 
&  $\frac 12(e_1 + e_4 + e_5)\,,$  \\
& $ \frac 12(e_3 + e_4)\,,\,  \frac 12 (e_1 + e_2)\,, $ 
& $\frac 12 (e_2 + e_4 + e_5)\,,$ \\
& $ \frac 12(e_2 + e_3)\,,\, \frac 12(e_1 + e_3)\,, $ 
& $\frac 12 (e_3 + e_4 + e_5)\,,$ \\
& $ \frac 12(e_1 + e_4) \,,\, \frac12(e_2 + e_4)\,,$ 
& $\frac 12 (e_1 + e_2 + e_3)\,,$ \\
& $ \frac 12 e_1 \,,\, \frac 12 e_2\,,\, \frac 12 e_3\,,\,\frac  12 e_4\,,\, \underline{0} $ 
& $\frac 12(e_1 + e_3)\,,\, \frac 12 (e_4 + e_5)\,,$\\
& & $\frac 12(e_1 + e_2)\,,\, \frac 12(e_2 + e_3)\,,$  \\
& & $\frac 12 e_1\,,\, \frac 12 e_2\,,\, \frac 12 e_3\,,\, \underline{0}$   
\\ \hline 
\end{tabular}
}
\end{center}
\end{table}
}
%

%
\newcommand{\ZeightIFixedPoints}{
\begin{figure}[ht!!] 
\caption{The four--tori are displayed on which the orbifold element $\gth^4$ acts non--trivially for the three $\mathbb{Z}_\text{8--I}$ compatible lattices. 
In these diagrams the vectors spanning $\mathbb{R}^4$ are indicated as the labels of the four coordinate axes. 
A colored region denotes the $\mathbb{R}^2$ plane spanned by the corresponding vectors with the indicated angle among them; a plane spanned by orthogonal basis vectors is greyed out. Only some characteristic fixed points  (fixed two--tori in $T^6$) of the $\theta^4$--sector are indicated for clarity. 
\label{Z8I_T4_torus}}  
\begin{center} 
\tabu{cc}{
\includegraphics[width=0.48\textwidth]{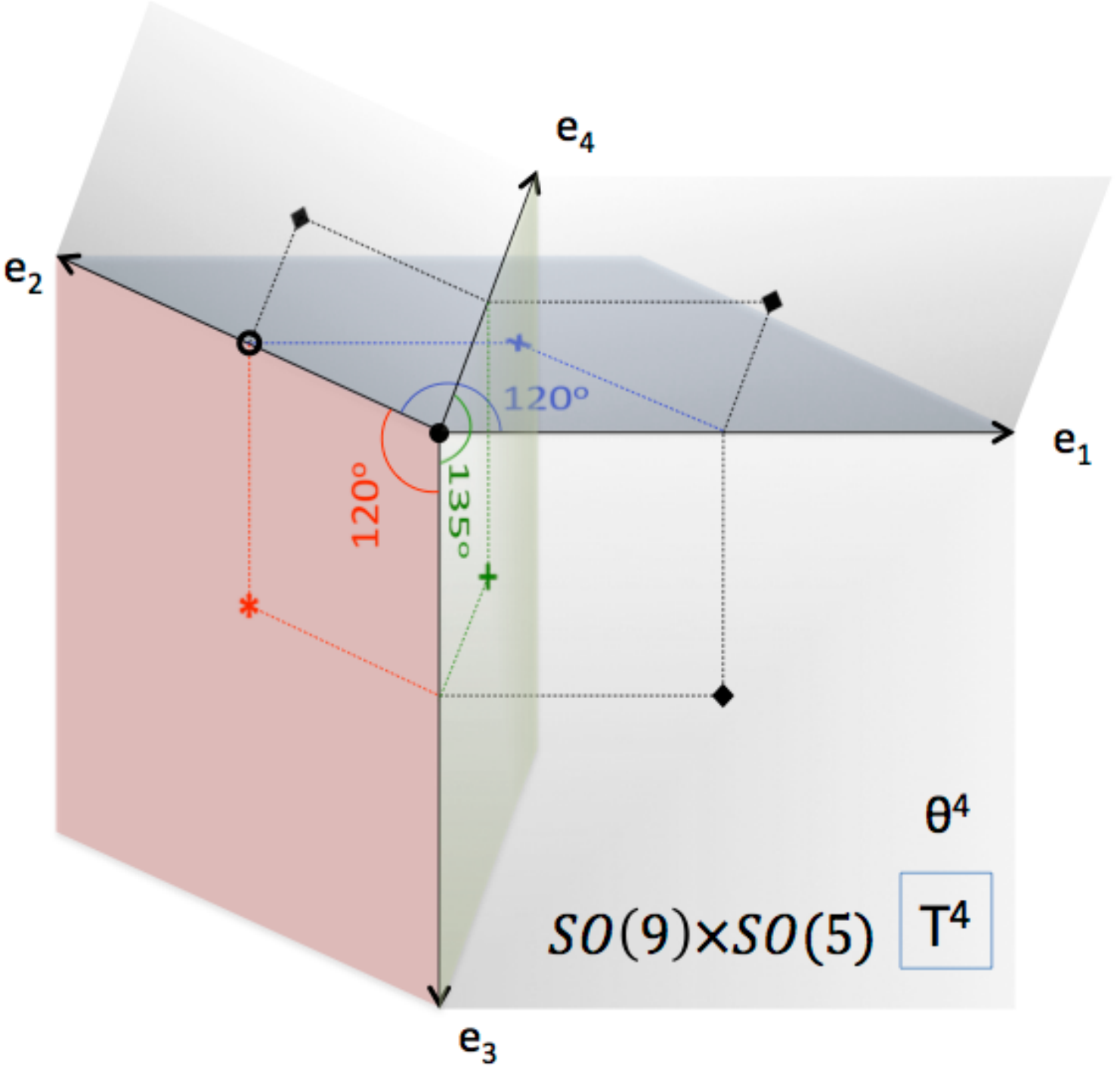}
& 
\includegraphics[width=0.48\textwidth]{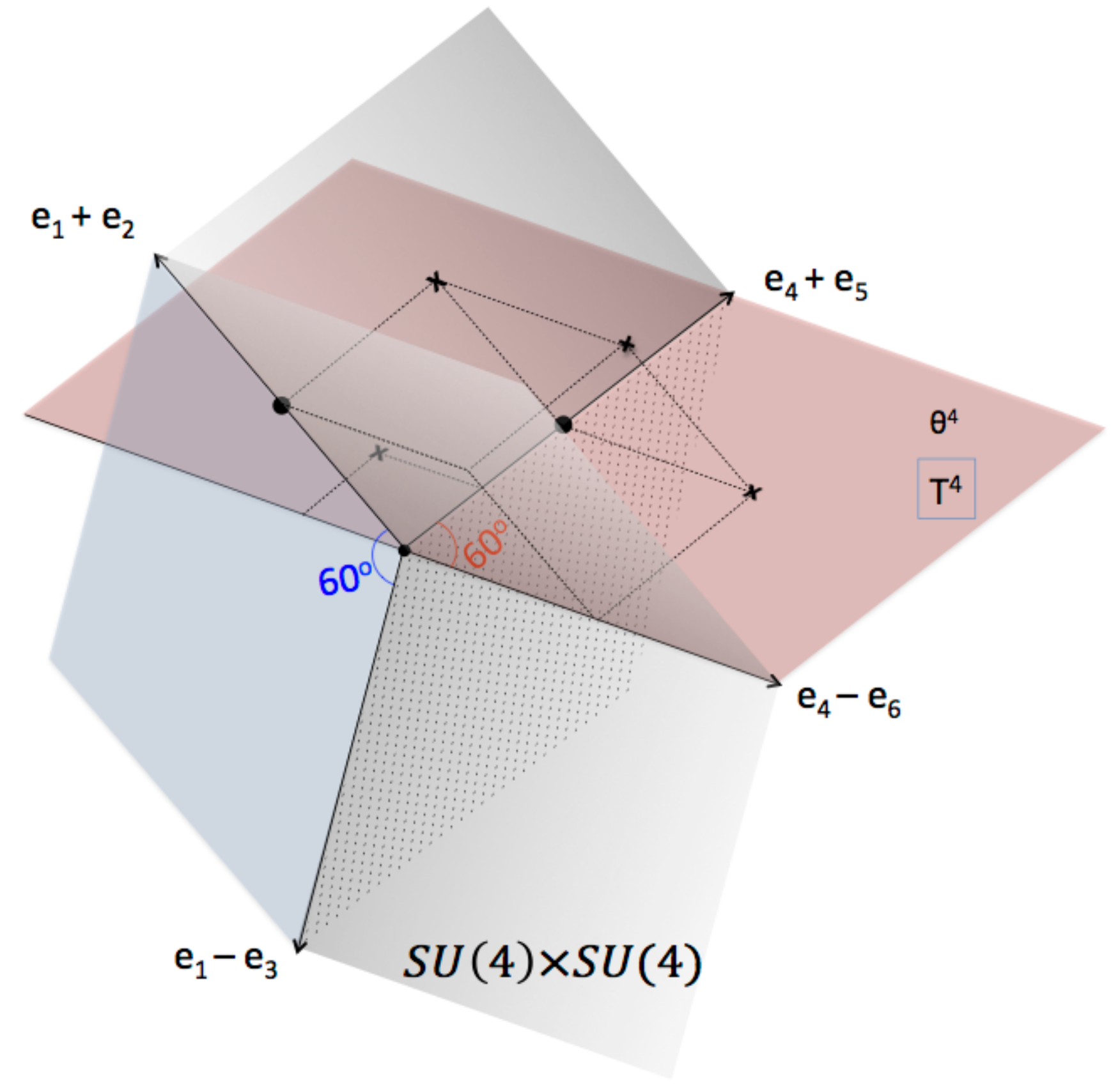}
\\[1ex] 
\multicolumn{2}{c}{
\includegraphics[width=0.48\textwidth]{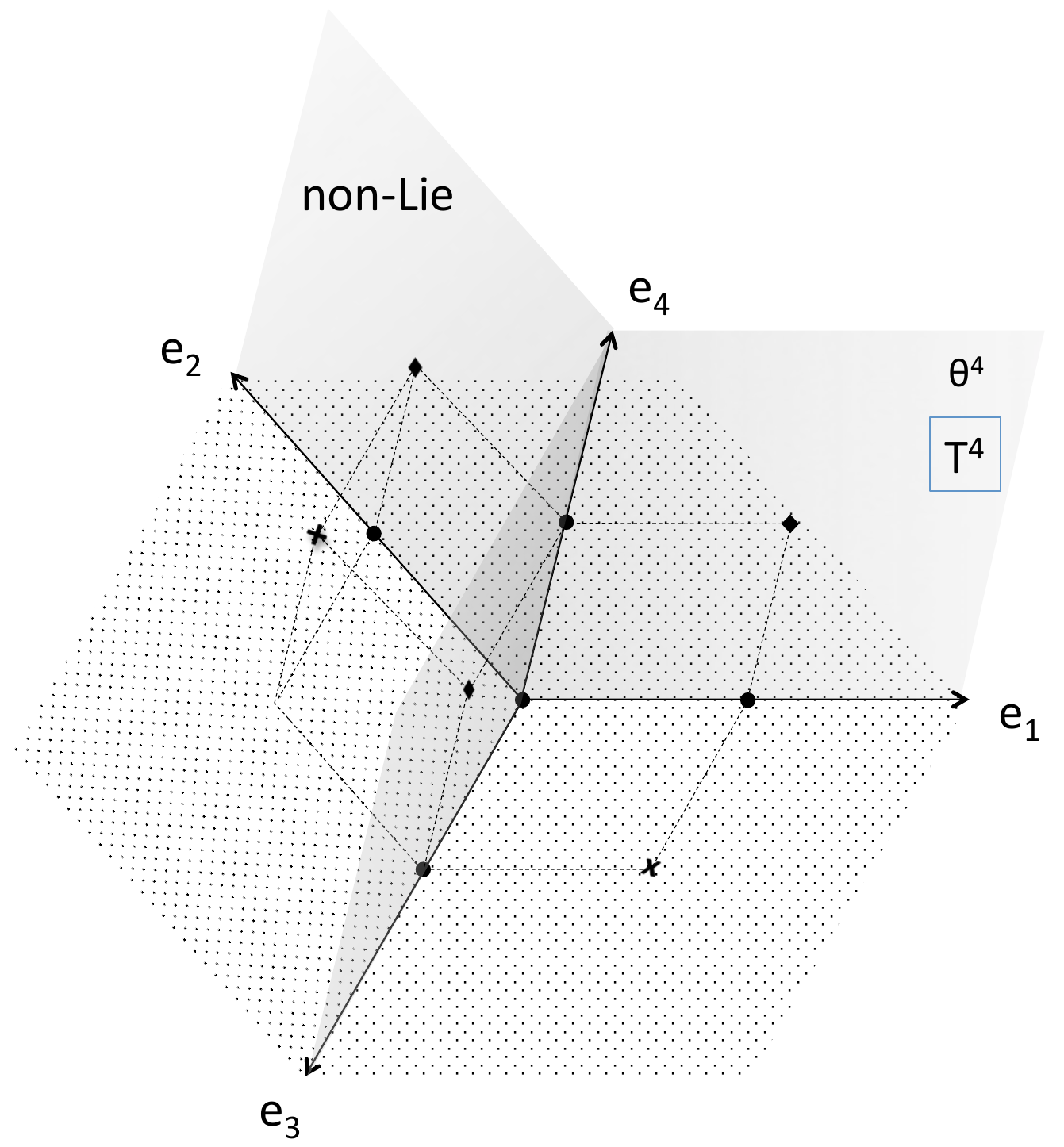}} 
}
\end{center} 
\end{figure}
}

%
\newcommand{\ZeightIIFixedPoints}{ 
\begin{figure}[ht!!] 
\caption{
The four--tori are displayed on which $\gth^2$ and $\gth^4$ act non--trivially for the two $\mathbb{Z}_\text{8--II}$ compatible lattices. 
The procedure followed to draw these diagrams is the same as in Figure~\ref{Z8I_T4_torus}. 
\label{Z8II_T4_torus}}  
\tabu{cc}{
\includegraphics[width=0.48\textwidth]{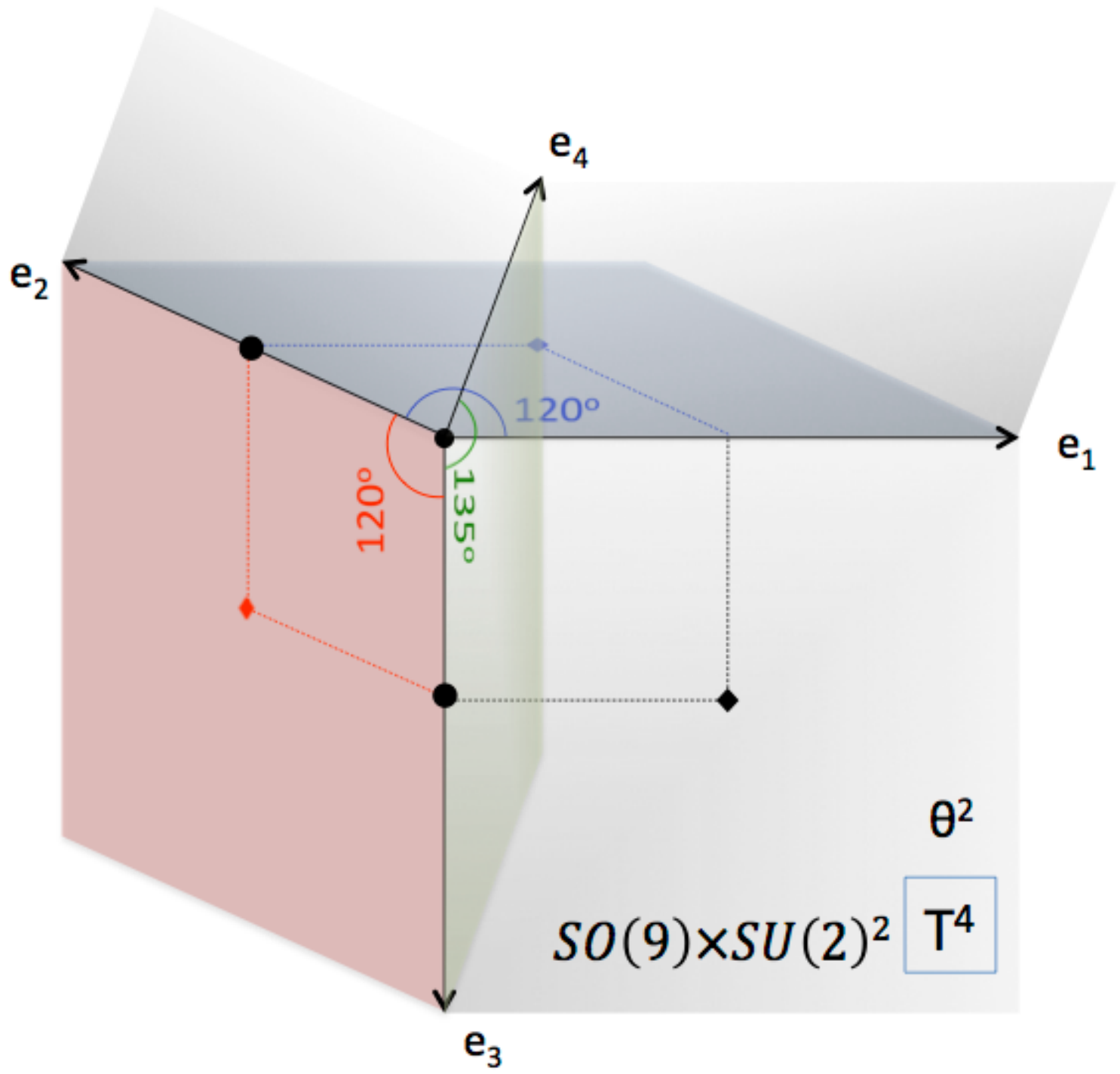}
& 
\includegraphics[width=0.48\textwidth]{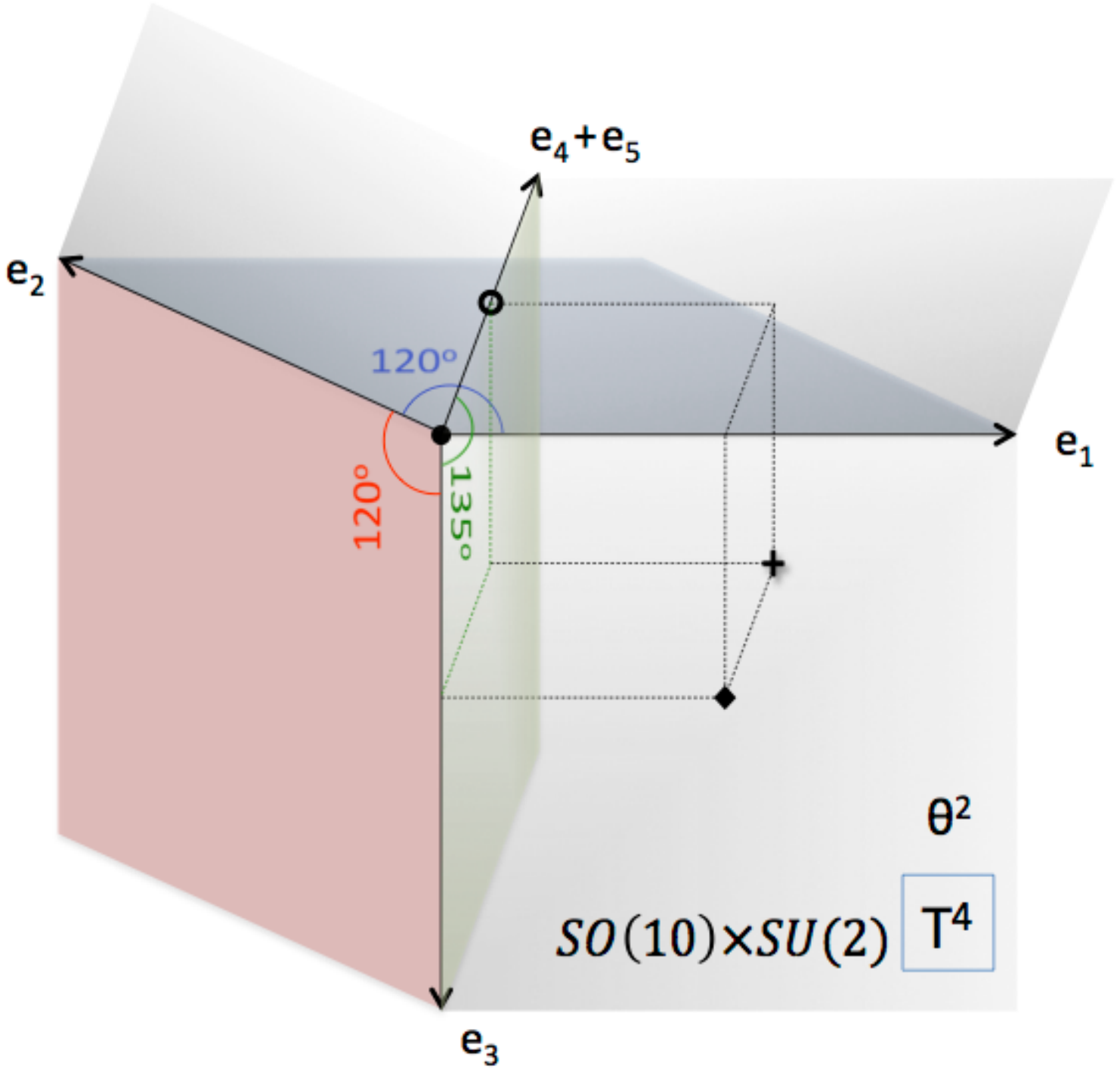}
\\[1ex] 
\includegraphics[width=0.48\textwidth]{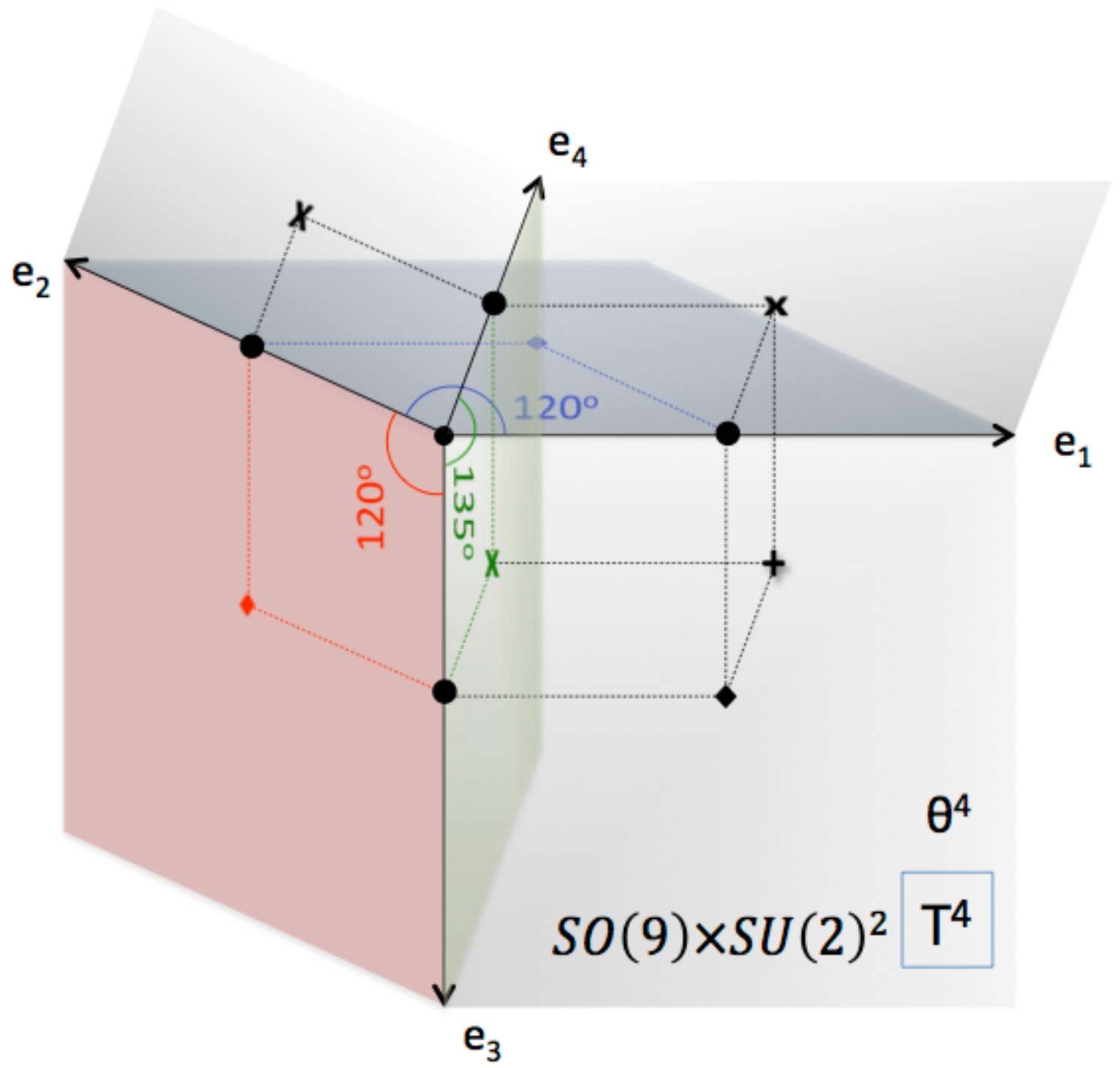}
& 
\includegraphics[width=0.48\textwidth]{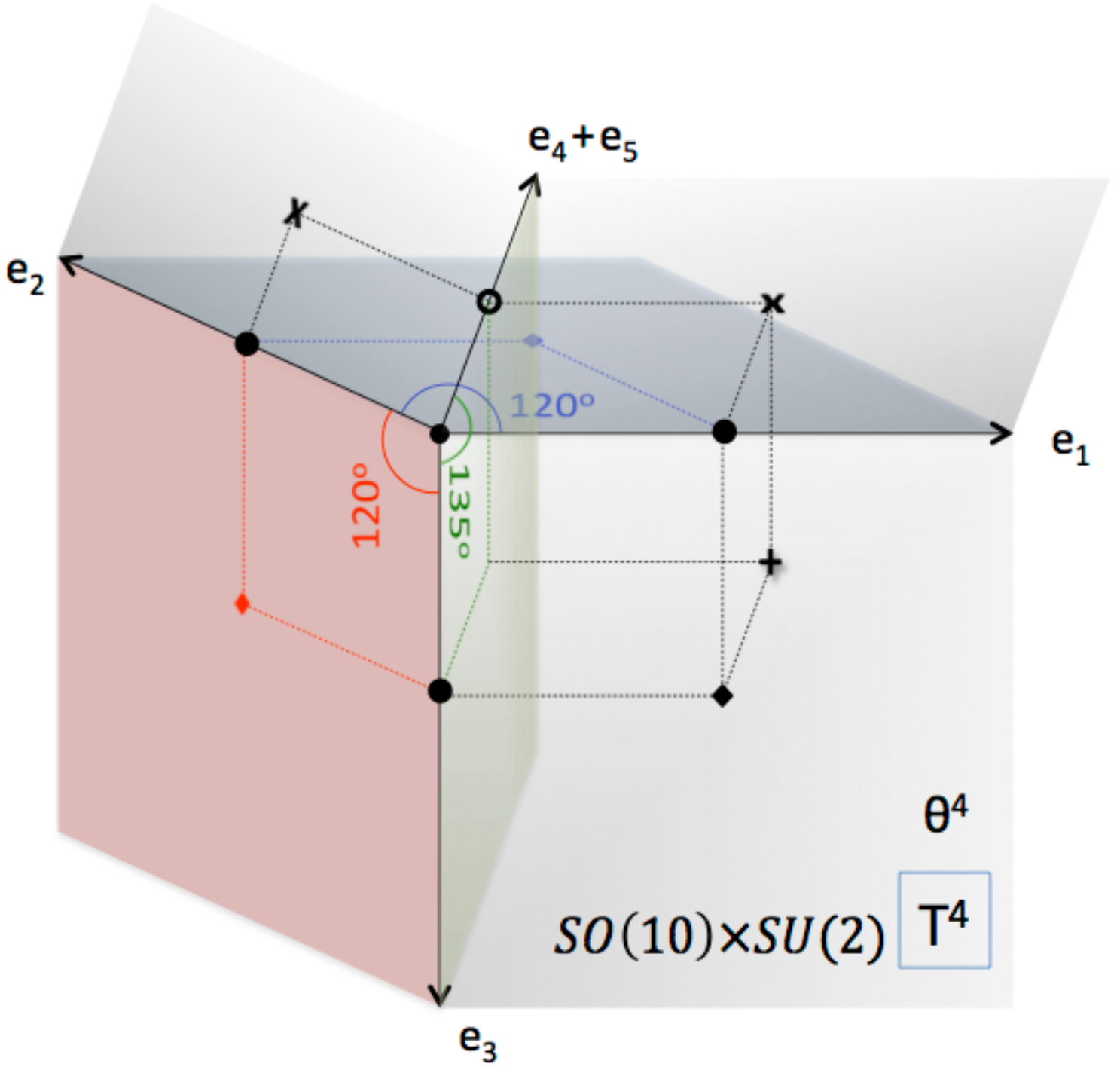}
}
\end{figure}
}

\section{Heterotic strings on $\Intr_8$ orbifolds}
\label{sc:Z8geoms} 

This section provides some specifics of $\Intr_8$ orbifolds.  In fact, it turns out that there are two inequivalent choices for the twist vector for such orbifolds. We refer to them as $\Intr_\text{8--I}$ and $\Intr_\text{8--II}$, respectively, which we define in detail below. Neither choice allows for a fully factorized lattice, therefore these all define non--factorizable orbifolds. Both of them are compatible with only a few inequivalent lattices. Moreover, one of them is not isomorphic to any Lie--lattice. In Tables~\ref{FixedPointsZ8I} and~\ref{FixedPointsZ8II} we give a comprehensive overview of the complete fixed sets of these orbifolds. 
In order to avoid over count states in the massless spectrum one should take into account that some fixed sets in the second and fourth twisted sectors come in conjugacy classes containing two or more elements. For the sake of brevity the conjugacy classes are not indicated in these Tables. 
Figures~\ref{Z8I_T4_torus} and~\ref{Z8II_T4_torus} provide schematic pictures of the positions of the fixed two--tori. In addition, we derive the consequences for the gauge shift and the discrete Wilson lines.

\clearpage
\ZeightIFixedSets
\clearpage
\ZeightIIFixedSets

\subsection{$\mathbf{\Intr_\text{8--I}}$  orbifolds}
\label{sc:Z8I}

In this subsection we describe some features of the $\Intr_\text{8--I}$ orbifolds. Their orbifold twist reads  
\begin{equation}  \label{Z8I twist vector}
\upsilon = \frac{1}{8}(1, -3, 2)~. 
\end{equation}
Consequently, the gauge shift vector has order 8: $8\, V \cong 0$. The $\gth^k$--twisted sectors, for $k=5,6,7$, offer no new information concerning the fixed point structures, since they are simply the anti--twists of the $k=3,2,1$ twisted sectors, respectively. Strings in the $\gth^4$--twisted sector can move freely over the fixed two--tori and thus define six--dimensional states. The $\Intr_\text{8--I}$ orbifold twist admits three inequivalent lattices. Two of them are the root lattices of $SO(9)\times SO(5)$ and $SU(4)\times SU(4)$ Lie--algebras, respectively, while the third one is a non--Lie lattice: 

\subsubsection*{$\mathbf{\Intr_\text{8--I}}$  orbifold on the $\mathbf{SO(9)\times SO(5)}$ lattice}

%
We take the torus lattice $\Gamma$ to be the root lattice of $SO(9) \times SO(5)$ with Dynkin diagrams:  
\[
\includegraphics[width=0.5\textwidth]{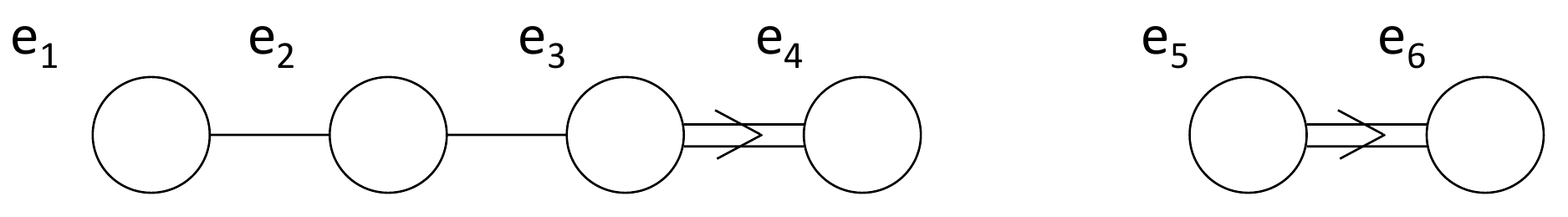}
\]
In the orthonormal basis of $\mathbb{R}^6$ we make the conventional choice \cite{liegroups} to express the basis vectors of this lattice as
\begin{equation} \label{root vectors Z8}
\begin{array}{lcrrrrrr}
e_1{}^T & = & (1, & -1, & ~~0, & ~~0, & ~~0, & ~~0)~,  
\\[1ex] 
e_2{}^T & = & (0, & 1, & -1, & 0, & 0, & 0)~,  
\\[1ex] 
e_3{}^T & = & (0, & 0, & 1, & -1, & 0, & 0)~,  
\\[1ex]
e_4{}^T & = & (0, & 0, & 0, & 1, & 0, & 0)~,  
\\[2ex] 
e_5{}^T & = & (0, & 0, & 0, & 0, & 1, & -1)~, 
\\[1ex] 
e_6{}^T & = & (0, & 0, & 0, & 0, & 0, & 1)~.  
\end{array}
\end{equation}
The Cartan metric of the $SO(9) \times SO(5)$ root lattice, 
\begin{equation}\label{metricSO9SO5}
G = 
\left( 
\begin{array}{rrrr|rr} 
2 & -1 & 0 & 0 & 0 & 0 \\
-1 & 2 & -1 & 0 & 0 & 0 \\
0 & -1 & 2 & -1 & 0 & 0 \\
0 & 0 & -1 & 1 & 0 & 0 \\ \hline 
0 & 0 & 0 & 0 & 2 & -1 \\
0 & 0 & 0 & 0 & -1 & 1  
\end{array}
\right)~, 
\end{equation}
can be computed using \eqref{torus metric}. 
%
%
The orbifold twist \eqref{Z8I twist vector} acts on this lattice basis by  
\begin{gather}  \label{TwistZ8ISO9SO5}
\arry{c}{ 
Q = 
\left( 
\begin{array}{rrrr|rr}
0 & 0 & 1 & -1 & 0 & 0 \\
1 & 0 & 1 & -1 & 0 & 0 \\
0 & 1 & 1 & -1 & 0 & 0 \\
0 & 0 & 2 & -1 & 0 & 0 \\ \hline 
0 & 0 & 0 & 0 & 1 & -1 \\
0 & 0 & 0 & 0 & 2 & -1 \\
\end{array}
\right)~. 
 }
\end{gather}
%
%
The action of the orbifold twist on the lattice vectors, given in \eqref{TwistZ8ISO9SO5},  implies that the discrete Wilson lines satisfy the following conditions  
\equ{
W_1 \cong W_2 \cong W_3 \cong 0~, 
\quad 
2\, W_4 \cong 0~, 
\qquad 
W_5 \cong 0~, 
\quad 
2\, W_6 \cong 0~.  
}
In other words only $W_4$ and $W_6$ constitute non--trivial discrete Wilson lines and they both have order $N_4=N_6 =2$.

%
Table \ref{FixedPointsZ8I} lists all fixed sets of the $\Intr_\text{8--I}$ orbifolds. The fixed points on $T^6$ for $\gth,\gth^2,\gth^3$--twisted sectors and the fixed tori of the $\gth^4$--sector are given in the second column of this Table for the $SO(9)\times SO(5)$ lattice. The effects of the $\gth$-- and $\gth^3$--twisted sectors turn out to be equivalent; they deliver the same fixed set and are therefore grouped together. The first twisted sector has four fixed points: From the direct product of the $\Intr_\text{8--I}$ orbifold they are given as two points that lie on the $T^2$ torus and two points within $T^4$.

%
Given that the fixed point structure for non--factorizable orbifolds is more involved than factorizable ones, we have developed the following schematic method to depict fixed sets in four dimensions, presented in Figure~\ref{Z8I_T4_torus}. This is particularly useful, when the six--torus can be partially factorized as $T^4 \times T^2$ or for sectors in which fixed two--tori are present. 
The first picture of Figure~\ref{Z8I_T4_torus} show the fundamental four--torus region of the $\Intr_\text{8--I}$ orbifold on the lattice $SO(9)\times SO(5)$ on which the orbifold element $\gth^4$ acts non--trivially. The labels of the four coordinate axes indicate the four lattice vectors spanning this four--torus. Pairs of two basis vectors span two--dimensional planes. When two such vectors are not perpendicular this plane is colored and their angle is indicated; planes spanned by orthogonal vectors are displayed in gray. Some $\gth^4$--fixed points are displayed in this Figure by dots, circles, etc. For example, the red star $*$ is located at $\frac 12(e_2+e_3)$ which is indeed one of the $\gth^4$--fixed points as can be seen from the second column of Table~\ref{FixedPointsZ8I}. For clarity we have refrained from indicating all possible fixed points in Figure~\ref{Z8I_T4_torus}.

\subsubsection*{$\mathbf{\Intr_\text{8--I}}$  orbifold on the $\mathbf{SU(4)\times SU(4)}$ lattice}

For the $SU(4)\times SU(4)$ Lie lattice we choose basis vectors
\begin{equation} \label{root vectors Z8-I_3}
\begin{array}{lc rrr rrr}
e_1{}^T & = & (\,~~\frac{1}{\sqrt{2}}, & 0, & 0, & -1, & -\frac{1}{\sqrt{2}}, & 0)~,  
\\[1ex] 
e_2{}^T & = & (\,~~~~0, & \frac{1}{\sqrt{2}}, & 0, & 1, & 0, & -\frac{1}{\sqrt{2}})~,  
\\[1ex]
e_3{}^T & = & (-\frac{1}{\sqrt{2}}, & 0, & 0, & -1, & \frac{1}{\sqrt{2}}, & 0)~, 
\\[2ex] 
e_4{}^T & = & (\,~~~~\frac{1}{2}, & \frac{1}{2}, & 1, & 0, & \frac{1}{2}, & \frac{1}{2})~,  
\\[1ex]
e_5{}^T & = & (~-\frac{1}{2}, & \frac{1}{2}, & -1, & 0, & -\frac{1}{2}, & \frac{1}{2})~,  
\\[1ex] 
e_6{}^T & = & (~-\frac{1}{2}, & -\frac{1}{2}, & 1, & 0, & -\frac{1}{2}, & -\frac{1}{2})~.  
\end{array}
\end{equation}
The Cartan metric of this Lie lattice computed using \eqref{torus metric} takes the form
\begin{equation}\label{metric_SU4xSU4}
G = 
\left( 
\begin{array}{rrr|rrr}
 2 & -1 &  0  &  0 &  0 &  0 \\
-1 &  2 & -1  &  0 &  0 &  0 \\
 0 & -1 &  2  &  0 &  0 &  0 \\ \hline 
 0 &  0 &  0  &  2 & -1 &  0 \\ 
 0 &  0 &  0  & -1 &  2 & -1 \\
 0 &  0 &  0  & 0 & -1 &  2 \\
\end{array}
\right)~.  
\end{equation}
The transformation properties of this basis under the $\mathbb{Z}_\text{8--I}$ action are described by the twist matrix
\begin{equation}\label{twist matrix Z8_I_3}
Q = 
\left( 
\begin{array}{rrr|rrr}
0 & ~~0 & ~~0  &~~0 &~~0 & -1 \\
0 & 0 & 0  &   1 & 0 & -1 \\
0 & 0 & 0  &   0 & 1 & -1 \\ \hline 
1 & 0 & 0  &  0 & 0 & 0 \\ 
0 & 1 & 0  &  0 & 0 & 0 \\
0 & 0 & 1  &  0 & 0 & 0 \\
\end{array}
\right)~.  
\end{equation}
The resulting fixed sets in the various orbifold sectors are given in the third column of Table~\ref{FixedPointsZ8I}. The position of the fixed two--tori in the fourth twisted sector are displayed in the second picture in Figure~\ref{Z8I_T4_torus}. 
From the basis vector transformation properties we determine the relations among Wilson lines,
\begin{equation} \label{WLsSU42} 
W_1 \approx W_2 \approx W_3 \approx W_4 \approx W_5 \approx W_6~, 
\quad
4\, W_6 \approx 0~,
\end{equation}
which means that we have six equivalent Wilson lines of order $N_6=4$.

\subsubsection*{$\mathbf{\Intr_\text{8--I}}$  orbifold on the non--Lie lattice}

Not all possible tori, compatible with $\mathcal{N}=1$ supersymmetry, can be spanned by Lie lattices. Following the classification of Ref.~\cite{nonlocal_patrick}, there is a non--Lie lattices for the $\mathbb{Z}_\text{8--I}$ orbifold, spanned by the lattice vectors 
\begin{equation} \label{root vectors Z8-I_2}
\begin{array}{lc rrr rrr}
e_1{}^T & = & (\,~~~~1, & ~0, & ~~~0, & ~~~0, & 1, & 0)~,  
\\[1ex] 
e_2{}^T & = &(\,~~\frac{1}{\sqrt{2}}, &  \frac{1}{\sqrt{2}}, & 0, & 0, & - \frac{1}{\sqrt{2}}, & - \frac{1}{\sqrt{2}})~,  
\\[1ex]
e_3{}^T & = & (\,~~~~0, & 1, & 0, & 0, & 0, & 1)~,  
\\[1ex]
e_4{}^T & = & (- \frac{1}{\sqrt{2}}, &  \frac{1}{\sqrt{2}}, & 0, & 0, &  \frac{1}{\sqrt{2}}, & - \frac{1}{\sqrt{2}})~,  
\\[2ex] 
e_5{}^T &=& \multicolumn{6}{c}{
\sfrac12  \big( e_1+e_2+e_3+e_4\big)^T + (0,0,1,-1,0,0)~,} 
\\[2ex]   
e_6{}^T  &=& \multicolumn{6}{c}{ \sfrac12  \big( e_1+e_2+e_3+e_4\big)^T + (0,0,1,~~1,0,0)~. }
\end{array} 
\end{equation}
The first four vectors are mutually orthonormal with norm 2. Because of the halfs in front of the sum of lattice vectors $e_1+\ldots + e_4$, the vectors $e_5$ and $e_6$ can never be transformed such that they have zero inner product with the basis vectors $e_1,\ldots, e_6$. Consequently, this constitutes a genuine non--Lie lattice. The Gram metric~\eqref{torus metric} of this non--Lie lattice is given by 
\begin{equation}\label{metric_nonLie2}
G = 
\left( 
\begin{array}{rrrr|rr} 
2 & 0 & 0 & 0 & 1 & 1 \\
0 & 2 & 0 & 0 & 1 & 1 \\
0 & 0 & 2 & 0 & 1 & 1 \\
0 & 0 & 0 & 2 & 1 & 1 \\ \hline 
1 & 1 & 1 & 1 & 4 & 2 \\
1 & 1 & 1 & 1 & 2 & 4  \\
\end{array}
\right)~.
\end{equation}

\clearpage 

\ZeightIFixedPoints

The twist matrix  in basis (\ref{root vectors Z8-I_2}) reads 
\begin{equation}\label{twist matrix Z8_I_2}
Q = 
\left( 
\begin{array}{rrrr|rr} 
 0 & ~~0 & ~~0 & -1  &  -1 &  0 \\
 1 & 0 & 0 &  0  &   0 &  1 \\
 0 & 1 & 0 &  0  &   0 &  1 \\
 0 & 0 & 1 &  0  &   0 &  1 \\ \hline 
 0 & 0 & 0 &  0  &   0 & -1 \\
 0 & 0 & 0 &  0  &   1 &  0 \\
\end{array}
\right)~. 
\end{equation}
The resulting fixed sets in the various orbifold sectors are given in the fourth column of Table~\ref{FixedPointsZ8I}. The position of the fixed two--tori in the fourth twisted sector are displayed in the third picture in Figure~\ref{Z8I_T4_torus}. From the transformation properties of basis system (\ref{twist matrix Z8_I_2}) we determine the relations between the discrete Wilson lines to be 
\begin{equation}
W_1 \cong W_2 \cong W_3 \cong W_4~, 
\quad 
2\, W_4 \cong 0~,  
\qquad 
W_5 \cong W_6 - W_4~, 
\quad 2\, W_6 \cong 0 ~,
\end{equation}
that is we have two inequivalent sets of Wilson lines both of order $N_4=N_6=2$.

\subsection{$\mathbf{\Intr_\text{8--II}}$ orbifolds}
\label{sc:Z8II}

In this subsection we describe some features of the $\Intr_\text{8--II}$ orbifolds. Their orbifold twist reads
\begin{equation}  \label{Z8II twist vector}
\upsilon = \frac{1}{8}(1, 3, -4)~, 
\end{equation}
so that the gauge shift vector has again order 8: $8\, V \cong 0$. 
As before, the $\gth^k$--twisted sectors for $k=5,6,7$ are simply the anti--twists of the $k=3,2,1$ twisted sectors. Contrary to the $\Intr_\text{8--I}$ orbifolds, now both the $\gth^2$-- and $\gth^4$--twisted sectors define six--dimensional states as they can propagate over fixed two--tori. The $\Intr_\text{8--II}$ orbifold action admits two inequivalent Lie lattices:

\subsubsection*{$\mathbf{\Intr_\text{8--II}}$  orbifold on the $\mathbf{SO(9)\times SU(2)^2}$ lattice}

Since the four--torus lattice of the $\Intr_\text{8--II}$ orbifold on the lattice $SO(9) \times SU(2)^2$ is identical to that of the $\Intr_\text{8--I}$ orbifold with lattice $SO(9)\times SO(5)$, we can recycle many of the properties discussed in Subsection~\ref{sc:Z8I}. This holds in particular for the Cartan metric and the generalized Coxeter matrix $Q$. We take the basis vectors of $\mathbb{R}^2$ to span $SU(2) \times SU(2)$, so that the Cartan metric becomes
\begin{equation}\label{metricSO9SU22}
G = 
\left( 
\begin{array}{rrrr|rr} 
2 & -1 & 0 & 0 & ~~0 & ~~0 \\
-1 & 2 & -1 & 0 & 0 & 0 \\
0 & -1 & 2 & -1 & 0 & 0 \\
0 & 0 & -1 & 1 & 0 & 0 \\ \hline 
0 & 0 & 0 & 0 & 2 & 0 \\
0 & 0 & 0 & 0 & 0 & 2  
\end{array}
\right)~. 
\end{equation}
In addition, the generalized Coxeter element reads 
\begin{equation} \label{TwistZ8IISO9SU22} 
Q = 
\left(
\begin{array}{rrrr|rr}
 0 & ~~0 & ~~1 & -1 & 0 & 0 \\
 1 & 0 & 1 & -1 & 0 & 0 \\
 0 & 1 & 1 & -1 & 0 & 0 \\
 0&  0 & 2 & -1& 0 & 0 \\ \hline
0 & 0 & 0 & 0 & -1 & 0 \\ 
0 & 0 & 0 & 0 & 0 & -1 \\
\end{array}
\right)~. 
\end{equation}
The fixed set configurations are summarized for the first four sectors in the second column of Table~\ref{FixedPointsZ8II}. The positions of some characteristic fixed two--tori in the second and fourth twisted sectors are displayed in the first column of Figure~\ref{Z8II_T4_torus}. The action of the orbifold twist on the lattice vectors, given in \eqref{TwistZ8IISO9SU22},  implies that the discrete Wilson lines satisfy the following conditions  
\equ{
W_1 \cong W_2 \cong W_3 \cong 0~, 
\quad 
2\, W_4 \cong 0~, 
\qquad 
2\, W_5 \cong 0~, 
\qquad 
2\, W_6 \cong 0~.  
}
Hence, there are three  non--trivial discrete Wilson lines $W_4, W_5$ and $W_6$ all of order $N_4=N_5=N_6 =2$.

\subsubsection*{$\mathbf{\Intr_\text{8--II}}$  orbifold on the $\mathbf{SO(10)\times SU(2)}$ lattice}

The $\Intr_\text{8--II}$ orbifold with lattice $SO(10) \times SU(2)$, with Dynkin diagram
\[
\includegraphics[width=0.5\textwidth]{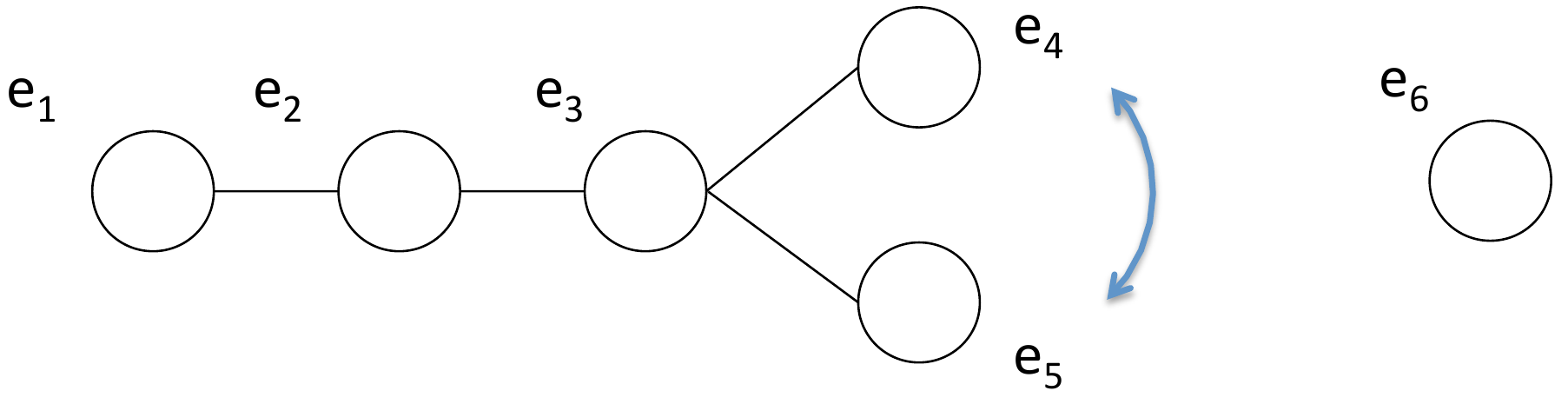}
\]
cannot be splitted as a $T^4\times T^2$. We make again the conventional choice of basis vectors for the underlying root lattice:
\begin{equation}\label{root vectors Z8-II_2}
\begin{array}{lc rrr rrr}
e_1{}^T & = & (1, & -1, & 0, & 0, & 0, & 0)~, 
\\[1ex] 
e_2{}^T & = & (0, & 1, & -1, & 0, & 0, & 0)~,  
\\[1ex] 
e_3{}^T & = & (0, & 0, & 1, & -1, & 0, & 0)~,  
\\[1ex] 
e_4{}^T & = & (0, & 0, & 0, & 1, & -1, & 0)~,  
\\[1ex] 
e_5{}^T & = & (0, & 0, & 0, & 1, & 1, & 0)~,  
\\[2ex] 
e_6{}^T & = & (0, & 0, & 0, & 0, & 0, & \sqrt{2})~. 
\end{array}
\end{equation}
The resulting Cartan metric reads
\equ{ 
G = 
\left( 
\begin{array}{rrrrr|r}
 2 & -1 &  0 &  0 &  0  & ~~0 \\
-1 &  2 & -1 &  0 &  0  &  0 \\
 0 & -1 &  2 & -1 & -1  &  0 \\ 
 0 &  0 & -1 &  2 &  0  &  0 \\
 0 &  0 & -1 &  0 &  2  &  0 \\ \hline 
 0 &  0 &  0 &  0 &  0  &  2
\end{array}
\right)~. 
}

\clearpage 
\ZeightIIFixedPoints

The Coxeter element in the simple root basis (\ref{root vectors Z8-II_2}) has then the following form:
\begin{equation} \label{TwistZ8IISO10SU2}
Q = 
\left( 
\begin{array}{rrrrr|r}
0 & ~~0 & ~~1 & -1 & -1 & 0 \\
1 & 0 & 1 & -1 & -1 & 0 \\
0 & 1 & 1 & -1 & -1 & 0 \\
0 & 0 & 1 & -1 & 0 & 0 \\
0 & 0 & 1 & 0 & -1 & 0 \\ \hline 
0 & 0 & 0 & 0 & 0 & -1 \\
\end{array} 
\right)~.
\end{equation}
In the third column of Table \ref{FixedPointsZ8II} the fixed point structure of $\mathbb{Z}_\text{8--II}$ on $SO(10) \times SU(2)$ is listed. The positions of some characteristic fixed two--tori in the second and fourth twisted sectors are displayed in the second column of Figure~\ref{Z8II_T4_torus}. 
%
%
The action of the orbifold twist on the lattice vectors, given in \eqref{TwistZ8IISO10SU2},  implies that the discrete Wilson lines satisfy the following conditions  
\equ{
W_1 \cong W_2 \cong W_3~, 
\quad 
W_3 \cong 2\, W_5~, 
\quad 
W_4 \cong - W_5~,
\quad 
4\, W_5 \cong 0~, 
\qquad 
2\, W_6 \cong 0~.  
}
This tells us that there are two independent non--trivial discrete Wilson lines $W_5$ and $W_6$ with orders $N_5 =4$ and $N_6=2$, respectively.  The Wilson lines $W_\ga$, $\ga=1,2,3$ and $W_4$ are all related to $W_5$.

%% file: models.tex
%
\newcommand{\ZeightSUShifts}{
\begin{table}[ht!!]
\caption{
The modular invariant $\mathbb{Z}_\text{8--I}$ and $\mathbb{Z}_\text{8--II}$ gauge shifts, that lead to an $SU(5)$ gauge group factor in the observable sector,  are listed. The complete semi--simple observable and hidden groups are given for every shift; the number of $U(1)$ factors is such that the total rank equals eight in both the observable and hidden groups. 
\label{shift_commulative}}
\hspace{-3ex}
\subtable[$\mathbb{Z}_\text{8--I}$ gauge shifts]
{\label{shift_commulative_I}  
\scalebox{.6}{
\renewcommand{\arraystretch}{1}
\begin{tabular}{|c||c|c|c|}
\hline 
\textbf{\#} & \textbf{Shift} & 
\multicolumn{2}{|c|}{\textbf{Gauge group}}
\\
& $\mathbf{8\, V}$ & \textbf{observable} & \textbf{hidden} 
\\ \hline \hline 
1 & 
(2 2   2  2  1  1  1 \!-1; 2  2  2  2  2  2  1  1)
&\multirow{7}{*}{$SU(5)\!\!\times\!\! SU(4)$} 
 & $E_6$\\ 
2 & 
(2  2  2  2  1  1  1 \!-1; 2  1  1  1  1  1  1  0)
& & $SO(12)$ \\ 
3 & 
(2  2  2  2  1  1  1 \!-1; 3  3  2  2  0  0  0  0) 
& & $SO(12)$ \\ 
4 & 
(2  2  2  2  1  1  1 \!-1; 3  3  3  3  2  1  1  0) 
& & $SO(8)\!\!\times\!\! SU(2)$\\ 
5 & 
(2  2  2  2  1  1  1 \!-1; 4  4  4  3  1  0  0  0)
& & $SU(8)$ \\ 
6 & 
(2  2  2  2  1  1  1 \!-1; 2  2  2  2  2  2  1 \!-1)
& & $SU(6)\!\!\times\!\! SU(2)$ \\ 
7 &
(2  2  2  2  1  1  1 \!-1; 3  3  2  2  2  2  2 \!-2)
& &$SU(6)\!\!\times\!\! SU(2)$ \\  \hline 
8 &
(3  3  3  3  2  2  2 \!-2; 3  3  2  2  2  2  2 \!-2)
& \multirow{7}{*}{$SU(5)\!\!\times\!\! SU(4)$}  
& $SU(6)\!\!\times\!\! SU(2)$ \\ 
9 & 
(3  3  3  3  2  2  2 \!-2; 2  2  2  2  2  2  1  1)
&  & $E_6$  \\ 
10 & 
(3  3  3  3  2  2  2 \!-2; 2  1  1  1  1  1  1  0)
& & $SO(12)$  \\  
11&
(3  3  3  3  2  2  2 \!-2; 3  3  2  2  0  0  0  0) 
& &  $SO(12)$  \\ 
12 & 
(3  3  3  3  2  2  2 \!-2; 3  3  3  3  2  1  1  0)
& &  $SO(8)\!\!\times\!\! SU(2)^2$  \\ 
13 & 
(3  3  3  3  2  2  2 \!-2; 4  4  4  3  1  0  0  0)
& & $SU(8)$  \\  
14 & 
(3  3  3  3  2  2  2 \!-2; 2  2  2  2  2  2  1 \!-1) 
& & $SU(6)\!\!\times\!\! SU(2)$  \\ \hline 
15 & 
(3  3  2  2  1  1  1 \!-1; 2  2  2  2  2  2  2  2)
& \multirow{7}{*}{$SU(5)\!\!\times\!\! SU(3)\!\!\times\!\! SU(2)$} 
& $E_7\!\!\times\!\! SU(2)$ \\ 
16 &
(3  3  2  2  1  1  1 \!-1; 8  0  0  0  0  0  0  0)
& & $SO(16)$ \\ 
17 & 
(3  3  2  2  1  1  1 \!-1; 2  2  2  2  0  0  0  0)
& & $SO(14)$ \\ 
18 & 
(3  3  2  2  1  1  1 \!-1; 3  3  3  3  3  1  1 \!-1) 
& & $SO(10)\!\!\times\!\! SU(4)$ \\  
19 & 
(3  3  2  2  1  1  1 \!-1; 2  2  2  2  2  2  2 \!-2) 
& & $SU(8)$ \\  
20 & 
(3  3  2  2  1  1  1 \!-1; 2  2  2  1  1  1  1  0)
& & $SU(7)$ \\ 
21 & 
(3  3  2  2  1  1  1 \!-1; 3  3  3  2  1  0  0  0)
& & $SU(7)$ \\ \hline 
22 & 
(3  3  3  3  2  2  1 \!-1; 2  2  2  2  2  2  2  2)
& \multirow{7}{*}{$SU(5)\!\!\times\!\! SU(3)\!\!\times\!\! SU(2)$}  
& $E_7\!\!\times\!\! SU(2)$ \\ 
23 &
(3  3  3  3  2  2  1 \!-1; 8  0  0  0  0  0  0  0)
& & $SO(16)$ \\ 
24 &
(3  3  3  3  2  2  1 \!-1; 2  2  2  2  0  0  0  0)
& & $SO(14)$ \\ 
25 & 
(3  3  3  3  2  2  1 \!-1; 3  3  3  3  3  1  1 \!-1)
& & $SO(10)\!\!\times\!\! SU(4)$ \\ 
26 & 
(3  3  3  3  2  2  1 \!-1; 2  2  2  2  2  2  2 \!-2) 
& & $SU(8)$ \\ 
27 & 
(3  3  3  3  2  2  1 \!-1; 2  2  2  1  1  1  1  0)
& & $SU(7)$ \\ 
28 & 
(3  3  3  3  2  2  1 \!-1; 3  3  3  2  1  0  0  0)
& & $SU(7)$ \\ \hline  
29 & 
(3  3  2  2  2  1  1  0; 3  3  3  3  2  2  1 \!-1)
& \multirow{7}{*}{$SU(5)\!\!\times\!\! SU(3)$} 
& $SU(5)\!\!\times\!\! SU(3)\!\!\times\!\! SU(2)$ \\ 
30 &
(3  3  2  2  2  1  1  0; 2  2  1  1  1  1  1  1)
& & $E_6$ \\  
31& 
(3  3  2  2  2  1  1  0; 3  2  2  2  2  2  1  0)
& & $SO(10)$ \\ 
32 & 
(3  3  2  2  2  1  1  0; 2  2  1  1  1  1  1 \!-1)
& & $SU(7)\!\!\times\!\! SU(2)$ \\ 
33 & 
(3  3  2  2  2  1  1  0; 3  3  3  3  3  3  2 \!-2)
& & $SU(7) \!\!\times\!\! SU(2)$ \\ 
34 & 
(3  3  2  2  2  1  1  0; 4  3  2  2  2  2  2 \!-1) 
& & $SU(6)\!\!\times\!\! SU(2)$ \\  
35 & 
(3  3  2  2  2  1  1  0; 3  3  2  2  1  1  1 \!-1)
& & $SU(5)\!\!\times\!\! SU(3)\!\!\times\!\! SU(2)$ \\ \hline   
36 & 
(3  3  3  2  2  1  1 \!-1; 1  1  1  1  1  1  1  1) 
& \multirow{9}{*}{$SU(5)\!\!\times\!\! SU(3)$} 
& $E_7$ \\ 
37 & 
(3  3  3  2  2  1  1 \!-1; 2  2  2  2  2  2  0  0) 
& & $E_6\!\!\times\!\! SU(2)$ \\  
38 &
(3  3  3  2  2  1  1 \!-1; 3  3  3  3  1  1  1  1)
& & $SO(12)\!\!\times\!\! SU(2)$ \\  
39 & 
(3  3  3  2  2  1  1 \!-1; 4  4  2  2  2  2  2 \!-2)
& & $SU(8) \times SU(2)$ \\ 
40 & 
(3  3  3  2  2  1  1 \!-1; 1  1  1  1  1  1  1 \!-1)
& & $SU(8)$ \\  
41 &
(3  3  3  2  2  1  1 \!-1; 4  4  4  2  1  1  1 \!-1)
& &  $SU(8)$ \\ 
42 & 
(3  3  3  2  2  1  1 \!-1; 3  2  2  2  1  1  1  0)
& & $SU(6)\!\!\times\!\! SU(2)$\\ 
43 & 
(3  3  3  2  2  1  1 \!-1; 3  3  3  2  2  2  1  0)
& & $SU(6)\!\!\times\!\! SU(2)$\\ 
44 &
(3  3  3  2  2  1  1 \!-1; 3  3  3  3  1  1  1 \!-1)
& & $SU(4)^2$ \\ \hline 
45 &
(4  3  3  2  2  2  1 \!-1; 2  2  1  1  1  1  1  1)
& \multirow{7}{*}{$SU(5)\!\!\times\!\! SU(3)$} 
& $E_6$ \\ 
46 & 
(4  3  3  2  2  2  1 \!-1; 3  2  2  2  2  2  1  0)
& & $SO(10)$ \\ 
47 & 
(4  3  3  2  2  2  1 \!-1; 2  2  1  1  1  1  1 \!-1)
& & $SU(7)\!\!\times\!\! SU(2)$ \\ 
48 & 
(4  3  3  2  2  2  1 \!-1; 3  3  3  3  3  3  2 \!-2)
& & $SU(7)\!\!\times\!\! SU(2)$\\  
49 &
(4  3  3  2  2  2  1 \!-1; 4  3  2  2  2  2  2 \!-1)
& & $SU(6) \!\!\times\!\! SU(2)$ \\ 
50 &
(4  3  3  2  2  2  1 \!-1; 3  3  2  2  1  1  1 \!-1)
& & $SU(5)\!\!\times\!\! SU(3)\!\!\times\!\! SU(2)$ \\  
51&
(4  3  3  2  2  2  1 \!-1; 3  3  3  3  2  2  1 \!-1)
& & $SU(5)\!\!\times\!\! SU(3) \!\!\times\!\! SU(2)$ \\ \hline  
52 & 
(3  3  2  2  2  2  1 \!-1; 2  2  2  2  2  2  1  1)
& \multirow{7}{*}{$SU(5)\!\!\times\!\! SU(2)^2$}  
& $E_6$ \\ 
53 & 
(3  3  2  2  2  2  1 \!-1; 2  1  1  1  1  1  1  0)
& & $SO(12)$ \\ 
54 & 
(3  3  2  2  2  2  1 \!-1; 3  3  2  2  0  0  0  0)
& & $SO(12)$ \\ 
55 &
(3  3  2  2  2  2  1 \!-1; 3  3  3  3  2  1  1  0)
& & $SO(8)\!\!\times\!\! SU(2)^2$ \\ 
56 & 
(3  3  2  2  2  2  1 \!-1; 4  4  4  3  1  0  0  0)
& & $SU(8)$ \\ 
57 & 
(3  3  2  2  2  2  1 \!-1; 2  2  2  2  2  2  1 \!-1)
& & $SU(6)\!\!\times\!\! SU(2)$ \\ 
58 & 
(3  3  2  2  2  2  1 \!-1; 3  3  2  2  2  2  2 \!-2)
& & $SU(6)\!\!\times\!\! SU(2)$ \\ \hline  
59 & 
(3  3  2  2  1  1  1 \!-1; 0  0  0  0  0  0  0  0)
& $SU(5)\!\!\times\!\! SU(3)\!\!\times\!\! SU(2)$ & $E_8$ \\ \hline 
60 & 
(3  3  3  3  2  2  1 \!-1; 0  0  0  0  0  0  0  0)
& $SU(5)\!\!\times\!\! SU(3)\!\!\times\!\! SU(2)$ & $E_8$ \\  \hline 
\end{tabular}
}}
\subtable[\label{shift_commulative_II}
$\mathbb{Z}_\text{8--II}$ gauge shifts]{
\scalebox{.6}{
\renewcommand{\arraystretch}{1}
\begin{tabular}{|c|c|c|c|}
\hline
\textbf{\#} & \textbf{Shift} & 
\multicolumn{2}{|c|}{\textbf{Gauge group}}
\\
& $\mathbf{8\, V}$ & \textbf{observable} & \textbf{hidden} 
\\ \hline \hline 
1 & 
(2  2  2  2  1  1  1 \!-1; 3  2  2  1  1  1  1 \!-1) 
& \multirow{7}{*}{$SU(5)\!\!\times\!\! SU(4)$}  
& $SU(7)$ \\
2 & 
(2  2  2  2  1  1  1 \!-1; 3  3  2  2  2  2  2  0)
& & $SU(6)\!\!\times\!\! SU(2)^2$ \\
3 & 
(2  2  2  2  1  1  1 \!-1; 4  3  3  2  2  2  2 \!-2)
& & $SU(6)\!\!\times\!\! SU(2)^2$ \\
4 & 
(2  2  2  2  1  1  1 \!-1; 1  1  1  1  1  1  0  0)
& & $E_6 \!\!\times\!\! SU(2)$ \\
5 & 
(2  2  2  2  1  1  1 \!-1; 3  3  3  3  3  3  0  0) 
& & $E_6\!\!\times\!\! SU(2)$ \\
6 & 
(2  2  2  2  1  1  1 \!-1; 2  2  2  2  2  1  1  0) 
& & $S0(10)\!\!\times\!\! SU(2)$ \\ 
7 &
(2  2  2  2  1  1  1 \!-1; 3  3  3  2  2  1  1 \!-1)
& & $SU(5)\!\!\times\!\! SU(3)$ \\ \hline 
8 & 
(3  3  3  3  2  2  2 \!-2; 3  2  2  1  1  1  1 \!-1)
& \multirow{5}{*}{$SU(5)\!\!\times\!\! SU(4)$}  
& $SU(7)$ \\
9 &
(3  3  3  3  2  2  2 \!-2; 3  3  2  2  2  2  2  0) 
& & $SU(6)\!\!\times\!\! SU(2)^2$ \\
10 & 
(3  3  3  3  2  2  2 \!-2; 4  3  3  2  2  2  2 \!-2)
&  & $SU(6)\!\!\times\!\! SU(2)^2$  \\
11 & 
(3  3  3  3  2  2  2 \!-2; 1  1  1  1  1  1  0  0) 
& & $E_6\!\!\times\!\! SU(2)$ \\
12 & 
(3  3  3  3  2  2  2 \!-2; 3  3  3  3  1  1  0  0)
&  & $SO(10)\!\!\times\!\! SU(2)$ \\ \hline 
13 &
(3  3  2  2  1  1  1 \!-1; 2  2  1  1  1  1  0  0)
& \multirow{5}{*}{$SU(5)\!\!\times\!\! SU(3)\!\!\times\!\! SU(2)$}   
& $SO(10)\!\!\times\!\! SU(3)$ \\
14 &
(3  3  2  2  1  1  1 \!-1; 3  3  3  3  2  2  0  0)
& & $SO(10)\!\!\times\!\! SU(3)$ \\ 
15 & 
(3  3  2  2  1  1  1 \!-1; 3  2  2  2  2  1  1 \!-1) 
& & $SO(8)\!\!\times\!\! SU(3)$ \\ 
16 & 
(3  3  2  2  1  1  1 \!-1; 2  2  2  2  2  2  2  0) 
&  & $SU(7)$\\ 
17 & 
(3  3  2  2  1  1  1 \!-1; 4  2  2  2  2  2  2 \!-2) 
&  & $SU(7)$ \\ \hline 
18 & 
(3  3  3  3  2  2  1 \!-1; 2  2  2  2  2  2  2  0) 
& \multirow{5}{*}{$SU(5)\!\!\times\!\! SU(3)\!\!\times\!\! SU(2)$}   
& $SU(7)$ \\
19 & 
(3  3  3  3  2  2  1 \!-1; 4  2  2  2  2  2  2 \!-2) 
&  & $SU(7)$ \\
20 & 
(3  3  3  3  2  2  1 \!-1; 3  2  2  2  2  1  1 \!-1) 
& & $SO(8)\!\!\times\!\! SU(3)$ \\ 
21 &
(3  3  3  3  2  2  1 \!-1; 3  3  3  3  2  2  0  0) 
& & $SO(10)\!\!\times\!\! SU(3)$ \\
22 &
(3  3  3  3  2  2  1 \!-1; 2  2  1  1  1  1  0  0) 
&  & $SO(10)\!\!\times\!\! SU(3)$ \\ \hline 
23 &
(3  3  2  2  2  1  1  0; 2  1  1  1  1  1  1  0)
& \multirow{4}{*}{$SU(5)\!\!\times\!\! SU(3)$}   
& $SO(12)$ \\
24 & 
(3  3  2  2  2  1  1  0; 3  3  2  2  0  0  0  0) 
&  & $SO(12)$ \\ 
25 & 
(3  3  2  2  2  1  1  0; 3  3  3  3  2  1  1  0) 
&  & $SO(8)\!\!\times\!\! SU(2)^2$ \\ 
26 &
(3  3  2  2  2  1  1  0; 4  4  4  3  1  0  0  0) 
&  & $SU(8)$ \\ \hline  
27 & 
(3  3  3  2  2  1  1 \!-1; 2  2  2  2  1  1  1 \!-1)
& \multirow{4}{*}{$SU(5)\!\!\times\!\! SU(3)$}   
& $SU(5)\!\!\times\!\! SU(4)$ \\
28 & 
(3  3  3  2  2  1  1 \!-1; 3  3  3  3  2  2  2 \!-2)
& & $SU(5)\!\!\times\!\! SU(4)$  \\
29 & 
(3  3  3  2  2  1  1 \!-1; 4  4  3  3  1  1  0  0) 
&  & $SO(8)\!\!\times\!\! SU(4)$ \\
30 & 
(3  3  3  2  2  1  1 \!-1; 3  3  2  2  2  2  1  1) 
&  & $SO(10)\!\!\times\!\! SU(2)^2$ \\ \hline 
31 &
(4  3  3  2  2  2  1 \!-1; 3  3  3  3  2  1  1  0)
& \multirow{7}{*}{$SU(5)\!\!\times\!\! SU(3)$}   
& $SO(8)\!\!\times\!\! SU(2)^2$ \\
32 &
(4  3  3  2  2  2  1 \!-1; 4  4  4  3  1  0  0  0)
& &  $SU(8)$ \\
33 & 
(4  3  3  2  2  2  1 \!-1; 2  2  2  2  2  2  1 \!-1)
&  & $SU(6)\!\!\times\!\! SU(2)$ \\
34 & 
(4  3  3  2  2  2  1 \!-1; 3  3  2  2  2  2  2 \!-2)
&  & $SU(6)\!\!\times\!\! SU(2)$ \\
35 & 
(4  3  3  2  2  2  1 \!-1; 3  3  2  2  0  0  0  0) 
&  & $SO(12)$ \\ 
36 & 
(4  3  3  2  2  2  1 \!-1; 2  2  2  2  2  2  1  1) 
&  &  $E_6$ \\
37 & 
(4  3  3  2  2  2  1 \!-1; 2  1  1  1  1  1  1  0) 
& &  $SO(12)$ \\ \hline  
38 &
(3  3  2  2  2  2  1 \!-1; 2  2  2  2  2  1  1  0)
& \multirow{8}{*}{$SU(5)\!\!\times\!\! SU(2)^2$}   
& $SO(10)\!\!\times\!\! SU(2)$ \\
39 & 
(3  3  2  2  2  2  1 \!-1; 3  3  3  3  1  1  0  0)
& & $SO(10)\!\!\times\!\! SU(2)$ \\
40 & 
(3  3  2  2  2  2  1 \!-1; 1  1  1  1  1  1  0  0)
&  & $E_6\!\!\times\!\! SU(2)$ \\ 
41 &
(3  3  2  2  2  2  1 \!-1; 3  3  3  3  3  3  0  0)
&  & $E_6 \!\!\times\!\! SU(2)$ \\
42 & 
(3  3  2  2  2  2  1 \!-1; 3  2  2  1  1  1  1 \!-1)
&  & $SU(7)$ \\
43 &
(3  3  2  2  2  2  1 \!-1; 3  3  2  2  2  2  2  0) 
&  & $SU(6)\!\!\times\!\! SU(2)^2$ \\
44 & 
(3  3  2  2  2  2  1 \!-1; 4  3  3  2  2  2  2 \!-2)
&  & $SU(6)\!\!\times\!\! SU(2)^2$ \\
45 &
(3  3  2  2  2  2  1 \!-1; 3  3  3  2  2  1  1 \!-1) 
& &  $SU(5)\!\!\times\!\! SU(3)$ 
\\ \hline 
\end{tabular} 
}}
\end{table}
}
%

%
\newcommand{\ZeightSOShifts}{
\begin{table}[ht!!]
\caption{
The modular invariant $\mathbb{Z}_\text{8--I}$ and $\mathbb{Z}_\text{8--II}$ gauge shifts, that lead to an $SO(10)$ gauge group factor in the observable sector,  are tabulated. 
\label{shift_commulative_SO}}
\hspace{-2ex}
\subtable[$\mathbb{Z}_\text{8--I}$ gauge shifts]
{\label{SO10shift_commulative_I}  
\scalebox{.6}{
\renewcommand{\arraystretch}{1}
\begin{tabular}{|c||c|c|c|}
\hline 
\textbf{\#} & \textbf{Shift} & 
\multicolumn{2}{|c|}{\textbf{Gauge group}}
\\
& $\mathbf{8\, V}$ & \textbf{observable} & \textbf{hidden} 
\\ \hline \hline 
1 &
(3  3  3  3  3  1  1 \!-1; 2  2  1  1  1  1  1  1)
&\multirow{5}{*}{$SO(10)\!\!\times\!\! SU(4)$} & $E_6$\\
2 &
(3  3  3  3  3  1  1 \!-1;  3  2  2  2  2  2  1  0) & & $SO(10)$\\
3 &
(3  3  3  3  3  1  1 \!-1;  2  2  1  1  1  1  1 \!-1) & & $SU(7)\!\!\times\!\! SU(2)$\\
4 &
(3  3  3  3  3  1  1 \!-1;  3  3  3  3  3  3  2 \!-2) & & $SU(7)\!\!\times\!\! SU(2)$\\
5 &
(3  3  3  3  3  1  1 \!-1; 4  3  2  2  2  2  2 \!-1) & & $SU(6)\!\!\times\!\! SU(2)$\\
\hline
6 &
(2  2  1  1  1  1  0  0; 1  1  0  0  0  0  0  0) & \multirow{8}{*}{$SO(10)\!\!\times\!\! SU(3)$} & $E_7$\\
7 &
(2  2  1  1  1  1  0  0; 3  3  0  0  0  0  0  0) & & $E_7$\\
8 &
(2  2  1  1  1  1  0  0; 3  2  2  2  2  2  2  1) & & $SO(12)\!\!\times\!\! SU(2)$\\
9 & 
(2  2  1  1  1  1  0  0; 4  4  3  3  0  0  0  0) & & $SO(12)\!\!\times\!\! SU(2)$\\
10 &
(2  2  1  1  1  1  0  0; 2  2  2  2  1  1  0  0) & & $SO(10)\!\!\times\!\! SU(2)$\\
11 &
(2  2  1  1  1  1  0  0; 3  3  2  2  2  2  0  0) & & $SO(10)\!\!\times\!\! SU(2)$\\
12 &
(2  2  1  1  1  1  0  0; 3  2  2  2  2  2  2 \!-1) & & $SU(6)$\\
13 &
(2  2  1  1  1  1  0  0; 4  3  3  3  2  1  1 \!-1) & & $SU(4)^2\!\!\times\!\! SU(2)$\\
\hline
14 &
(3  3  3  3  2  2  0  0; 1  1  0  0  0  0  0  0) & \multirow{8}{*}{$SO(10)\!\!\times\!\! SU(3)$} & $E_7$\\
15 &
(3  3  3  3  2  2  0  0; 3  3  0  0  0  0  0  0) & & $E_7$\\
16 &
(3  3  3  3  2  2  0  0; 3  2  2  2  2  2  2  1)  & & $SO(12)\!\!\times\!\! SU(2)$\\
17 & 
(3  3  3  3  2  2  0  0; 4  4  3  3  0  0  0  0)  & & $SO(12)\!\!\times\!\! SU(2)$\\
18 &
(3  3  3  3  2  2  0  0; 2  2  2  2  1  1  0  0) & & $SO(10)\!\!\times\!\! SU(2)$\\
19 &
(3  3  3  3  2  2  0  0; 3  3  2  2  2  2  0  0) & & $SO(10)\!\!\times\!\! SU(2)$\\
20 &
(3  3  3  3  2  2  0  0; 3  2  2  2  2  2  2 \!-1) & & $SU(6)$\\
21 &
(3  3  3  3  2  2  0  0; 4  3  3  3  2  1  1 \!-1) & & $SU(4)^2\!\!\times\!\! SU(2)$\\
\hline
22 &
(3  3  2  2  2  2  1  1; 3  3  3  3  2  1  1  0) & \multirow{7}{*}{$SO(10)\!\!\times\!\! SU(2)^2$} & $SO(8)\!\!\times\!\! SU(2)^2$\\
23 &
(3  3  2  2  2  2  1  1; 4  4  4  3  1  0  0  0) & & $SU(8)$\\
24 &
(3  3  2  2  2  2  1  1; 2  2  2  2  2  2  1 \!-1) & & $SU(6)\!\!\times\!\! SU(2)$\\
25 &
(3  3  2  2  2  2  1  1; 3  3  2  2  2  2  2 \!-2) & & $SU(6)\!\!\times\!\! SU(2)$\\
26 &
(2  2  2  2  2  2  1  1; 3  3  2  2  2  2  1  1) & & $SO(10)\!\!\times\!\! SU(2)^2$\\
27 &
(3  3  2  2  2  2  1  1; 2  1  1  1  1  1  1  0) & & $SO(12)$\\
28 &
(3  3  2  2  2  2  1  1; 3  3  2  2  0  0  0  0) & & $SO(12)$\\
\hline
29 &
(2  2  2  2  1  1  0  0;  3  2  2  2  2  1  1 \!-1) & \multirow{4}{*}{$SO(10)\!\!\times\!\! SU(2)$} & $SO(8)\!\!\times\!\! SU(3)$\\
30 &
(2  2  2  2  1  1  0  0; 2  2  2  2  2  2  2  0) & & $SU(7)$\\
31 &
(2  2  2  2  1  1  0  0; 4  2  2  2  2  2  2 \!-2) & & $SU(7)$\\
32 &
(2  2  2  2  1  1  0  0; 3  3  3  2  2  2  2 \!-1) & & $SU(4)\!\!\times\!\! SU(3)$\\
\hline
33 &
(2  2  2  2  2  1  1  0; 4  4  2  2  2  2  2 \!-2) & \multirow{9}{*}{$SO(10)\!\!\times\!\! SU(2)$} & $SU(8)\!\!\times\!\! SU(2)$\\
34 &
(2  2  2  2  2  1  1  0; 1  1  1  1  1  1  1 \!-1) & & $SU(8)$\\
35 &
(2  2  2  2  2  1  1  0; 4  4  4  2  1  1  1 \!-1) & & $SU(8)$\\
36 &
(2  2  2  2  2  1  1  0; 3  2  2  2  1  1  1  0) & & $SU(6)\!\!\times\!\! SU(2)$\\
37 &
(2  2  2  2  2  1  1  0; 3  3  3  2  2  2  1  0) & & $SU(6)\!\!\times\!\! SU(2)$\\
38 &
(2  2  2  2  2  1  1  0; 3  3  3  3  1  1  1 \!-1) & & $SU(4)^2$\\
39 &
(2  2  2  2  2  1  1  0; 1  1  1  1  1  1  1  1) & & $E_7$\\
40 & 
(2  2  2  2  2  1  1  0; 2  2  2  2  2  2  0  0) & & $E_6\!\!\times\!\! SU(2)$\\
41 &
(2  2  2  2  2  1  1  0; 3  3  3  3  1  1  1  1) & & $SO(12)\!\!\times\!\! SU(2)$\\
\hline
42 &
(3  3  2  2  2  2  0  0; 3  2  2  2  2  1  1 \!-1) & \multirow{4}{*}{$SO(10)\!\!\times\!\! SU(2)$} & $SO(8)\!\!\times\!\! SU(3)$\\
43 &
(3  3  2  2  2  2  0  0; 2  2  2  2  2  2  2  0) & & $SU(7)$\\
44 &
(3  3  2  2  2  2  0  0; 4  2  2  2  2  2  2 \!-2) & & $SU(7)$\\
45 &
(3  3  2  2  2  2  0  0; 3  3  3  2  2  2  2 \!-1) & & $SU(4) \times SU(3)$\\
\hline
46 &
(3  3  3  3  1  1  0  0; 4  4  2  2  2  2  2 \!-2) & \multirow{9}{*}{$SO(10)\!\!\times\!\! SU(2)$} & $SU(8)\!\!\times\!\! SU(2)$\\
47 &
(3  3  3  3  1  1  0  0; 1  1  1  1  1  1  1 \!-1) & & $SU(8)$\\
48 &
(3  3  3  3  1  1  0  0; 4  4  4  2  1  1  1 \!-1) & & $SU(8)$\\
49 &
(3  3  3  3  1  1  0  0; 3  2  2  2  1  1  1  0) & & $SU(6)\!\!\times\!\! SU(2)$\\
50 &
(3  3  3  3  1  1  0  0; 3  3  3  2  2  2  1  0) & & $SU(6)\!\!\times\!\! SU(2)$\\
51 &
(3  3  3  3  1  1  0  0; 3  3  3  3  1  1  1 \!-1) & & $SU(4)^2$\\
52 &
(3  3  3  3  1  1  0  0; 1  1  1  1  1  1  1  1) & & $E_7$\\
53 &
(3  3  3  3  1  1  0  0; 2  2  2  2  2  2  0  0) & & $E_6\!\!\times\!\! SU(2)$\\
54 &  
(3  3  3  3  1  1  0  0; 3  3  3  3  1  1  1  1) & & $SO(12)\!\!\times\!\! SU(2)$\\
\hline
55 &  
(3  2  2  2  2  2  1  0; 2  2  2  2  2  2  2 \!-2) & \multirow{6}{*}{$SO(10)$} & $SU(8)$\\
56 &
(3  2  2  2  2  2  1  0; 2  2  2  1  1  1  1  0) & & $SU(7)$\\
57 &
(3  2  2  2  2  2  1  0; 3  3  3  2  1  0  0  0) & & $SU(7)$\\
58 &
(3  2  2  2  2  2  1  0; 2  2  2  2  2  2  2  2) & & $E_7\!\!\times\!\! SU(2)$\\
59 & 
(3  2  2  2  2  2  1  0; 8  0  0  0  0  0  0  0) & & $SO(16)$\\
60 &
(3  2  2  2  2  2  1  0; 2  2  2  2  0  0  0  0) & & $SO(14)$
\\  \hline 
\end{tabular}
}}
\subtable[\label{SO10shift_commulative_II}
$\mathbb{Z}_\text{8--II}$ gauge shifts]{
\scalebox{.6}{
\renewcommand{\arraystretch}{1}
\begin{tabular}{|c|c|c|c|}
\hline
\textbf{\#} & \textbf{Shift} & 
\multicolumn{2}{|c|}{\textbf{Gauge group}}
\\
& $\mathbf{8\, V}$ & \textbf{observable} & \textbf{hidden} 
\\ \hline \hline 
1     & (3  3  3  3  3  1  1 \!-1; 2  2  2  2  2  2  1  1) & \multirow{7}{*}{$SO(10)\!\!\times\!\! SU(4)$} & $E_6$\\
2     & (3  3  3  3  3  1  1 \!-1; 2  1  1  1  1  1  1  0)  & & $SO(12)$\\
3     & (3  3  3  3  3  1  1 \!-1; 3  3  2  2  0  0  0  0)  & & $SO(12)$\\
4     & (3  3  3  3  3  1  1 \!-1;  3  3  3  3  2  1  1  0) & & $SO(8)\!\!\times\!\! SU(2)^2$\\
5     & (3  3  3  3  3  1  1 \!-1;  4  4  4  3  1  0  0  0) & & $SU(8)$\\
6     & (3  3  3  3  3  1  1 \!-1;  2  2  2  2  2  2  1 \!-1) & & $SU(6)\!\!\times\!\! SU(2)$\\
7     & (3  3  3  3  3  1  1 \!-1;  3  3  2  2  2  2  2 \!-2) & & $SU(6)\!\!\times\!\! SU(2)$\\
\hline
8     & (2  2  1  1  1  1  0  0;  3  2  2  2  2  2  1  0) & \multirow{7}{*}{$SO(10)\!\!\times\!\! SU(3)$} & $SO(10)$\\
9     & (2  2  1  1  1  1  0  0;  2  2  1  1  1  1  1 \!-1) & & $SU(7)\!\!\times\!\! SU(2)$\\
10    & (2  2  1  1  1  1  0  0;  3  3  3  3  3  3  2 \!-2) & & $SU(7)\!\!\times\!\! SU(2)$\\
11    & (2  2  1  1  1  1  0  0;  4  3  2  2  2  2  2 \!-1) & & $SU(6)\!\!\times\!\! SU(2)$\\
12    & (2  2  1  1  1  1  0  0;  3  3  2  2  1  1  1 \!-1) & & $SU(5)\!\!\times\!\! SU(3)\!\!\times\!\! SU(2)$\\
13    & (2  2  1  1  1  1  0  0;  3  3  3  3  2  2  1 \!-1) & & $SU(5)\!\!\times\!\! SU(3)\!\!\times\!\! SU(2)$\\
14    & (2  2  1  1  1  1  0  0;  2  2  1  1  1  1  1  1) & & $E_6$\\
\hline
15    & (3  3  3  3  2  2  0  0;  3  2  2  2  2  2  1  0) & \multirow{7}{*}{$SO(10)\!\!\times\!\! SU(3)$} & $SO(10)$\\
16    & (3  3  3  3  2  2  0  0;  2  2  1  1  1  1  1 \!-1) & & $SU(7)\!\!\times\!\! SU(2)$\\
17    & (3  3  3  3  2  2  0  0;  3  3  3  3  3  3  2 \!-2) & & $SU(7)\!\!\times\!\! SU(2)$\\
18    & (3  3  3  3  2  2  0  0;  4  3  2  2  2  2  2 \!-1) & & $SU(6)\!\!\times\!\! SU(2)$\\
19    & (3  3  3  3  2  2  0  0;  3  3  2  2  1  1  1 \!-1) & & $SU(5)\!\!\times\!\! SU(3)\!\!\times\!\! SU(2)$\\
20    & (3  3  3  3  2  2  0  0;  3  3  3  3  2  2  1 \!-1) & & $SU(5)\!\!\times\!\! SU(3)\!\!\times\!\! SU(2)$\\
21    & (3  3  3  3  2  2  0  0; 2  2  1  1  1  1  1  1)  & & $E_6$\\
\hline
22    & (3  3  2  2  2  2  1  1;  2  2  2  2  2  1  1  0) & \multirow{8}{*}{$SO(10)\!\!\times\!\! SU(2)^2$} & $SO(10)\!\!\times\!\! SU(2)$\\
23    & (3  3  2  2  2  2  1  1;  3  3  3  3  1  1  0  0) & & $SO(10)\!\!\times\!\! SU(2)$\\
24    & (3  3  2  2  2  2  1  1;  3  2  2  1  1  1  1 \!-1) & & $SU(7)$\\
25    & (3  3  2  2  2  2  1  1;  3  3  2  2  2  2  2  0) & & $SU(6)\!\!\times\!\! SU(2)^2$\\
26    & (3  3  2  2  2  2  1  1;  4  3  3  2  2  2  2 \!-2) & & $SU(6)\!\!\times\!\! SU(2)^2$\\
27    & (3  3  2  2  2  2  1  1;  3  3  3  2  2  1  1 \!-1) & & $SU(5)\!\!\times\!\! SU(3)$\\
28    & (3  3  2  2  2  2  1  1; 1  1  1  1  1  1  0  0)  & & $E_6\!\!\times\!\! SU(2)$\\
29    & (3  3  2  2  2  2  1  1; 3  3  3  3  3  3  0  0)  & & $E_6\!\!\times\!\! SU(2)$\\
\hline
30    & (2  2  2  2  1  1  0  0;  4  4  2  2  2  2  2 \!-2) & \multirow{7}{*}{$SO(10)\!\!\times\!\! SU(2)$} & $SU(8)\!\!\times\!\! SU(2)$\\
31    & (2  2  2  2  1  1  0  0;  1  1  1  1  1  1  1 \!-1) & & $SU(8)$\\
32    & (2  2  2  2  1  1  0  0;  4  4  4  2  1  1  1 \!-1) & & $SU(8)$\\
33    & (2  2  2  2  1  1  0  0;  3  2  2  2  1  1  1  0) & & $SU(6)\!\!\times\!\! SU(2)$\\
34    & (2  2  2  2  1  1  0  0;  3  3  3  2  2  2  1  0) & & $SU(6)\!\!\times\!\! SU(2)$\\
35    & (2  2  2  2  1  1  0  0;  3  3  3  3  1  1  1 \!-1) & & $SU(4)^2$\\
36    & (2  2  2  2  1  1  0  0; 3  3  3  3  1  1  1  1)  & & $SO(12)\!\!\times\!\! SU(2)$\\
\hline
37    & (2  2  2  2  2  1  1  0;  4  4  3  3  1  1  0  0)  & \multirow{7}{*}{$SO(10)\!\!\times\!\! SU(2)$} & $SO(8)\!\!\times\!\! SU(4)$\\
38    & (2  2  2  2  2  1  1  0;  2  2  2  2  1  1  1 \!-1) & & $SU(5)\!\!\times\!\! SU(4)$\\
39    & (2  2  2  2  2  1  1  0;  3  3  3  3  2  2  2 \!-2) & & $SU(5)\!\!\times\!\! SU(4)$\\
40    & (2  2  2  2  2  1  1  0;  3  3  2  2  2  2  1 \!-1) & & $SU(5)\!\!\times\!\! SU(2)^2$\\
41    & (2  2  2  2  2  1  1  0; 1  1  1  1  0  0  0  0)  & & $SO(14)$\\
42    & (2  2  2  2  2  1  1  0; 3  3  3  3  0  0  0  0)  & & $SO(14)$\\
43    & (2  2  2  2  2  1  1  0; 2  2  2  2  1  1  1  1)  & & $SO(12)$\\
\hline
44    & (3  3  2  2  2  2  0  0;  4  4  2  2  2  2  2 \!-2) & \multirow{9}{*}{$SO(10)\!\!\times\!\! SU(2)$} & $SU(8)\!\!\times\!\! SU(2)$\\
45    & (3  3  2  2  2  2  0  0;  1  1  1  1  1  1  1 \!-1) & & $SU(8)$\\
46    & (3  3  2  2  2  2  0  0;  4  4  4  2  1  1  1 \!-1) & & $SU(8)$\\
47    & (3  3  2  2  2  2  0  0;  3  2  2  2  1  1  1  0) & & $SU(6)\!\!\times\!\! SU(2)$\\
48    & (3  3  2  2  2  2  0  0;  3  3  3  2  2  2  1  0) & & $SU(6)\!\!\times\!\! SU(2)$\\
49    & (3  3  2  2  2  2  0  0;  3  3  3  3  1  1  1 \!-1) & & $SU(4)^2$\\
50    & (3  3  2  2  2  2  0  0; 1  1  1  1  1  1  1  1)  & & $E_7$\\
51    & (3  3  2  2  2  2  0  0; 2  2  2  2  2  2  0  0)  & & $E_7\!\!\times\!\! SU(2)$\\
52    & (3  3  2  2  2  2  0  0; 3  3  3  3  1  1  1  1)  & & $SO(12)\!\!\times\!\! SU(2)$\\
\hline
53    & (3  3  3  3  1  1  0  0;  4  4  3  3  1  1  0  0) & \multirow{7}{*}{$SO(10)\!\!\times\!\! SU(2)$} & $SO(8)\!\!\times\!\! SU(4)$\\
54    & (3  3  3  3  1  1  0  0;  2  2  2  2  1  1  1 \!-1) & & $SU(5)\!\!\times\!\! SU(4)$\\
55    & (3  3  3  3  1  1  0  0;  3  3  3  3  2  2  2 \!-2) & & $SU(5)\!\!\times\!\! SU(4)$\\
56    & (3  3  3  3  1  1  0  0;  3  3  2  2  2  2  1 \!-1) & & $SU(5)\!\!\times\!\! SU(2)^2$\\
57    & (3  3  3  3  1  1  0  0; 1  1  1  1  0  0  0  0)  & & $SO(14)$\\
58    & (3  3  3  3  1  1  0  0; 3  3  3  3  0  0  0  0)  & & $SO(14)$\\
59    & (3  3  3  3  1  1  0  0; 2  2  2  2  1  1  1  1)  & & $SO(12)$\\
\hline
60    & (3  2  2  2  2  2  1  0;  3  2  2  2  2  1  1 \!-1) & \multirow{4}{*}{$SO(10)$} & $SO(8)\!\!\times\!\! SU(3)$\\
61    & (3  2  2  2  2  2  1  0;  2  2  2  2  2  2  2  0) & & $SU(7)$\\
62    & (3  2  2  2  2  2  1  0;  4  2  2  2  2  2  2 \!-2) & & $SU(7)$\\
63    & (3  2  2  2  2  2  1  0;  3  3  3  2  2  2  2 \!-1) & & $SU(4)\!\!\times\!\! SU(3)$
\\ \hline 
\end{tabular} 
}}
\end{table}
}
%

%
\newcommand{\MSSMsummary}{
\begin{table}[ht!!]  
\renewcommand{\arraystretch}{1.2}
\caption{This table provides an overview of the MSSM spectra found on the $\Intr_8$ orbifolds based on the $SU(5)$ and $SO(10)$ shifts. The generation numbers are averaged over all independent models for each lattice separately.}\label{MSSM_averages_SUSO}
%
\begin{center}
\scalebox{.8}{
\begin{tabular}{l}
%
\begin{tabular}{|cc||c|c|c|c|c|c|c|c|c|c|}
 \hline
&& 
\multicolumn{10}{|c|}{\textbf{MSSM states and vector--like exotic pairs}} 
\\ 
\multicolumn{2}{|c||}{$\mathbf{\Intr_8}$ \textbf{Orbifolds}} & $q_i$ & $\overline{q}_i$ & $\overline{u}_i$ & $u_i$ & $\overline{d}_i$ & $d_i$ & $l_i$ & $h_i$ & $\overline{e}_i$ & $e_i$ 
\\ 
& & $(3,2)_{\frac 16}$ & $(\overline{3},2)_{-\frac 16}$ & $(\overline{3},1)_{-\frac 23}$ & $(3,1)_{\frac 23}$ &  $(\overline{3},1)_{\frac 13}$ & $(3,1)_{-\frac 13}$ & $(1,2)_{-\frac 12}$ & $(1,2)_{\frac 12}$ & $(1,1)_{1}$ & $(1,1)_{-1}$ \\
\hline \hline 
\multirow{3}{*}{$\Intr_\text{8--I}$} 
& $SO(9) \times SO(5)$ & 3.07 & 0.07 & 3.18 & 0.18 & 9.72 & 6.72 & 9.59 & 6.59 & 3.18 & 0.18 \\
\cline{2-12} 
&$SU(4) \times SU(4)$ & 3 & 0 & 3 & 0 & 6.08 & 3.08 & 5.90 & 2.90 & 3 & 0 \\
\cline{2-12} 
&non--Lie & 3.04 & 0.04 & 3.10 & 0.10 & 6.78 & 3.78 & 6.58 & 3.58 & 3.10 & 0.10 
\\ \hline\hline
\multirow{2}{*}{$\Intr_\text{8--II}$}  
& $SO(9) \times SU(2)^2$ & 3.25  & 0.25  & 3.08  & 0.08  & 8.16  & 5.16  & 7.53  & 4.53  & 3.08  & 0.08 \\
\cline{3-12}
& $SO(10) \times SU(2)$ & 3.21 & 0.21 & 3.05 & 0.05 & 7.37 & 4.37 & 5.99 & 2.99 & 3.05 & 0.05 \\
\hline
\end{tabular}
\\[-3ex] \\
%
%
\begin{tabular}{|cc||c|c|c|c|c|c|c|c||c|}
\hline 
&& \multicolumn{8}{|c||}{\textbf{Other vector--like exotic states}}
& \textbf{Total number}   
\\
\multicolumn{2}{|c||}{$\mathbf{\Intr_8}$ \textbf{Orbifolds}} & $s^+_i$ & $s^-_i$ & $s^0_i$ & $\phi_i$ & $\overline{\phi}_i$ & $w_i$ & $y_i$ & $\overline{y}_i$ 
& \,\textbf{of independent}\\
& & $(1,1)_{\frac 12}$ & $(1,1)_{-\frac 12}$ & $(1,1)_0$ & $(3,1)_{\frac 16}$ &  $(\overline{3},1)_{-\frac 16}$ & $(1,2)_0$ & $(3,2)_{\frac 13}$ & $(\overline{3},2)_{-\frac 13}$ 
& \textbf{models}
\\\hline \hline 
\multirow{3}{*}{$\Intr_\text{8--I}$}  & $SO(9) \times SO(5)$ & 23.63 & 23.63 & 122.48 & 3.53 & 3.53  & 10.34 & 0.02 & 0.02 & 180 \\
\cline{2-11} 
& $SU(4) \times SU(4)$ & 22.71 & 22.71 & 107.10 & 3.24 & 3.24 & 9.02 & 0 & 0 & 49 \\
\cline{2-11} 
& non-Lie & 22.62 & 22.62 & 104.99 & 3.26 & 3.26 & 9.56 & 0.01  & 0.01 & 81 
\\\hline\hline
\multirow{2}{*}{$\Intr_\text{8--II}$}  & $SO(9) \times SU(2)^2$ & 27.90 & 27.90 & 118.52 & 3.55  & 3.55  & 11.40 & 0.02  & 0.02 & 365 \\
\cline{2-11} 
& $SO(10) \times SU(2)$ & 14.33 & 14.33 & 95.82 & 1.77 & 1.77 & 5.79 & 0 & 0 & 78 \\
\hline
\end{tabular}
\end{tabular} 
}
\end{center}
\end{table}
}
%

%
\newcommand{\ExactThreeGen}{
\begin{table}[hb!!]
\caption{This Table gives the models with exactly three generations of quark--doublets. These models are classified into two categories whether their three quark--doublets are untwisted or twisted states. 
\label{Quarks_summary_SUSO}}  
\vspace{-1ex} 
\begin{center}
\begin{tabular}{|cc||c|c|c|c|c|}
\hline
\multicolumn{2}{|c||}{\multirow{2}{*}{$\mathbf{\Intr_8}$ \textbf{Orbifolds}}} & \textbf{Models with} & \multicolumn{4}{c|}{\textbf{Models with} $\mathbf{(N_{un}, N_\text{tw})}$} \\
&  & \textbf{exactly} $\mathbf{3\,q_i}$ & $\mathbf{(3,0)}$ & $\mathbf{(2,1)}$ & $\mathbf{(1,2)}$ & $\mathbf{(0,3)}$ \\
\hline \hline 
\multirow{3}{*}{$\mathbb{Z}_\text{8--I}$} & $SO(9) \times SO(5)$ &  174 & 81 & 22 & 71 & 0 
\\ \cline{2-7} 
& $SU(4) \times SU(4)$ & 49 & 45 & 4 & 0 & 0 
\\ \cline{2-7} 
& non-Lie & 78 & 37 & 8 & 33 & 0 
\\\hline\hline
\multirow{2}{*}{$\mathbb{Z}_\text{8--II}$} & $SO(9) \times SU(2)^2$ & 294 & 8 & 0 & 286 & 0 
\\ \cline{2-7} 
& $SO(10) \times SU(2)$ & 60 & 2 & 12 & 45 & 1 \\
\hline
\end{tabular}
\end{center}
\end{table}}
%

%
\newcommand{\HiddenGauge}{
\begin{table}[ht!!]
\caption{This Table gives the number of $\Intr_8$ orbifold models with a particular unbroken hidden gauge group.} \label{hidden_sectors_SUSO}
\vspace{-1ex}
\begin{center}
\scalebox{.9}{
\begin{tabular}{|c||c|c|c||c|c|}
\hline
\textbf{hidden sector} & \multicolumn{3}{|c||}{$\mathbf{\mathbb{Z}_\text{8--I}}$ \textbf{orbifold}} & \multicolumn{2}{|c|}{$\mathbf{\mathbb{Z}_\text{8--II}}$ \textbf{orbifold}} \\
\textbf{gauge group} & \multicolumn{1}{|c|}{$\mathbf{SO(9) \times SO(5)}$} & $\mathbf{SU(4) \times SU(4)}$ & \textbf{non--Lie} & $\mathbf{SO(9) \times SU(2)^2}$ & $\mathbf{SO(10) \times SU(2)}$ 
\\\hline\hline 
$SU(8)$ & 2 & 0 & 1 & 8 & 0 \\ 
$SU(7)$ & 0     & 1     & 0     & 0     & 0 \\
$SU(6) \times SU(2)$ & 1     & 1     & 1     & 12    & 3 \\
$SU(6)$ & 3     & 1     & 2     & 39    & 4 \\
$SO(12)$ & 0     & 0     & 0     & 0     & 1 \\
$SO(10) \times SU(2)^2$ & 0     & 0     & 0     & 0     & 1 \\
$SO(10) \times SU(2)$ & 0     & 1     & 0     & 0     & 1 \\
$SO(10)$ & 22    & 7     & 8     & 0     & 3 \\
$SU(5) \times SU(3)$ & 0     & 1     & 0     & 0     & 1 \\
$SU(5) \times SU(2)$ & 19    & 2     & 8     & 39    & 3 \\
$SU(5)$ & 19    & 2     & 11    & 55    & 9 \\
$SO(8) \times SU(4)$ & 11    & 0     & 5     & 0     & 0 \\
$SO(8) \times SU(2)^2$ & 8     & 0     & 3     & 0     & 0 \\
$SO(8)$ & 26    & 8     & 10    & 0     & 2 \\
$SU(4)^2$ & 2     & 0     & 1     & 12    & 3 \\
$SU(4) \times SU(2)^2$ & 17    & 5     & 5     & 21    & 4 \\
$SU(4) \times SU(2)$ & 5     & 3     & 4     & 49    & 14 \\
$SU(4)$ & 4 & 4 & 2 & 26 & 8 \\
$SU(3)^2 \times SU(2)$ & 10    & 1     & 5     & 33    & 6 \\
$SU(3) \times SU(3)$ & 9     & 0     & 4     & 11    & 1 \\
$SU(3) \times SU(2)^2$ & 9     & 4     & 4     & 16    & 5 \\
$SU(3) \times SU(2)$ & 0     & 3     & 1     & 27    & 6 \\
$SU(3)$ & 2     & 1     & 1     & 0     & 0 \\
$SU(2)^6$ & 7     & 0     & 3     & 0     & 0 \\
$SU(2)^4$ & 2     & 3     & 1     & 10    & 0 \\
$SU(2)^3$ & 2     & 1     & 1     & 0     & 1 \\
$SU(2)^2$ & 0     & 0     & 0     & 7     & 2 \\
\hline
\end{tabular}
}
\end{center}
\end{table}
}
%

%
\newcommand{\DistrQuarks}{
\begin{figure}[ht!!]
\caption{These histograms present the average (anti--)quark--doublets distributions over the different sectors for each orbifold. Here, the quark--doublets, $q_i$, are plotted in blue whereas the anti--quark--doublets, $\overline{q}_i$, in red. Along the $x$--axis we set out the orbifold sectors while on the $y$--axis the number of (anti--)quark--doublets.
\label{Quark_distribution_SUSO}}  
\hspace{-1ex} 
\tabu{cc}{
\tabu{c}{
(a) $\mathbb{Z}_\text{8--I}$ models
\\[1ex] 
\includegraphics[width=0.45\textwidth]{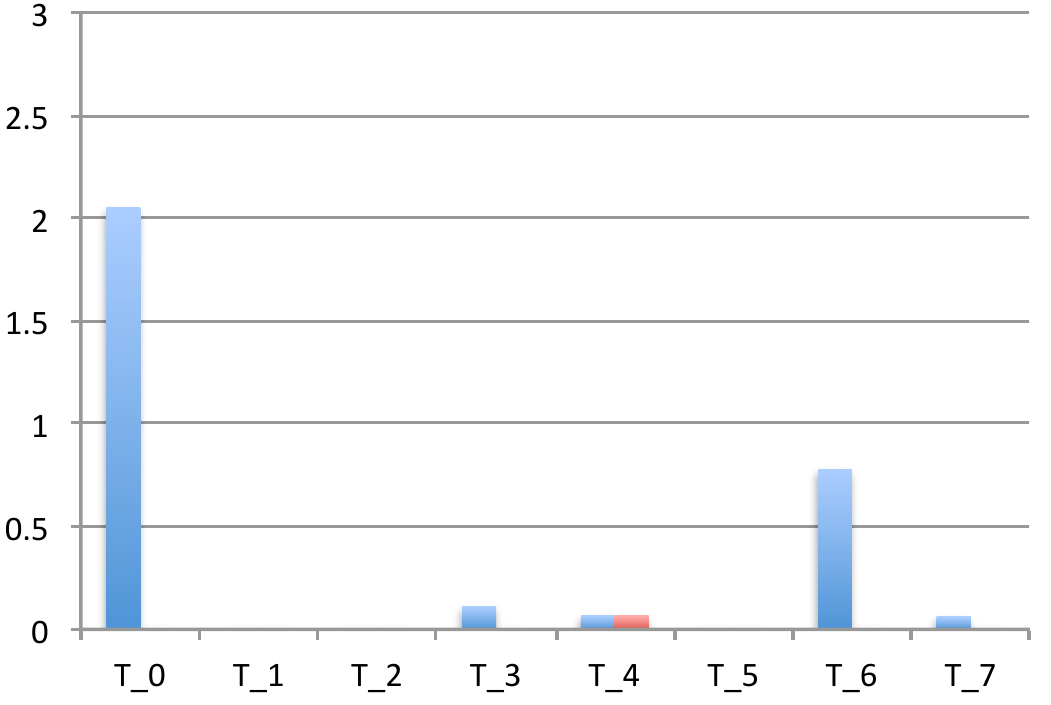}
\\
$SO(9)\times SO(5)$ lattice  
\\[1ex] 
\includegraphics[width=0.45\textwidth]{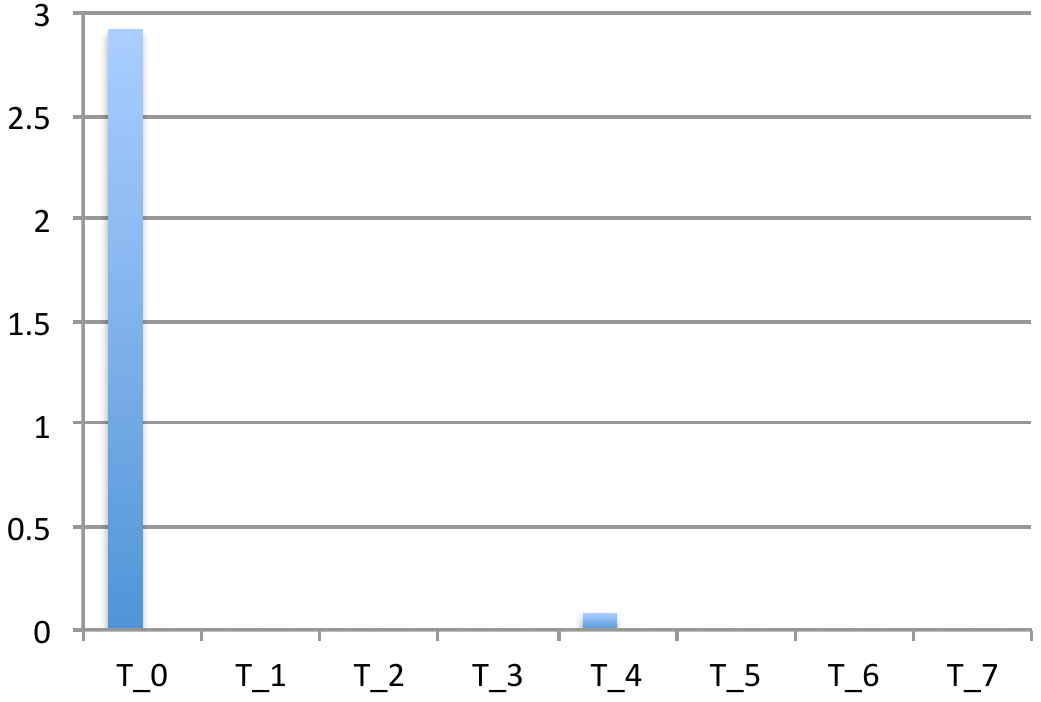}
\\
$SU(4)\times SU(4)$ lattice
\\[1ex]  
\hspace{.5mm} 
\includegraphics[width=0.45\textwidth]{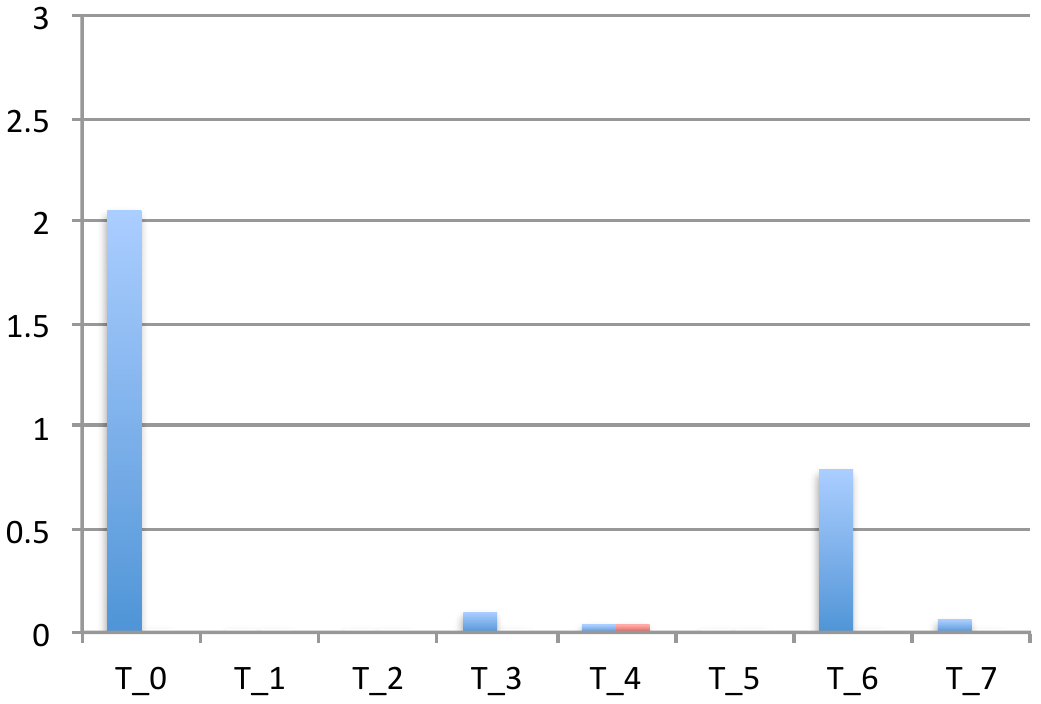}
\\
non--Lie lattice 
}
\tabu{c}{
(b) $\mathbb{Z}_\text{8--II}$ models
\\[1ex] 
\includegraphics[width=0.45\textwidth]{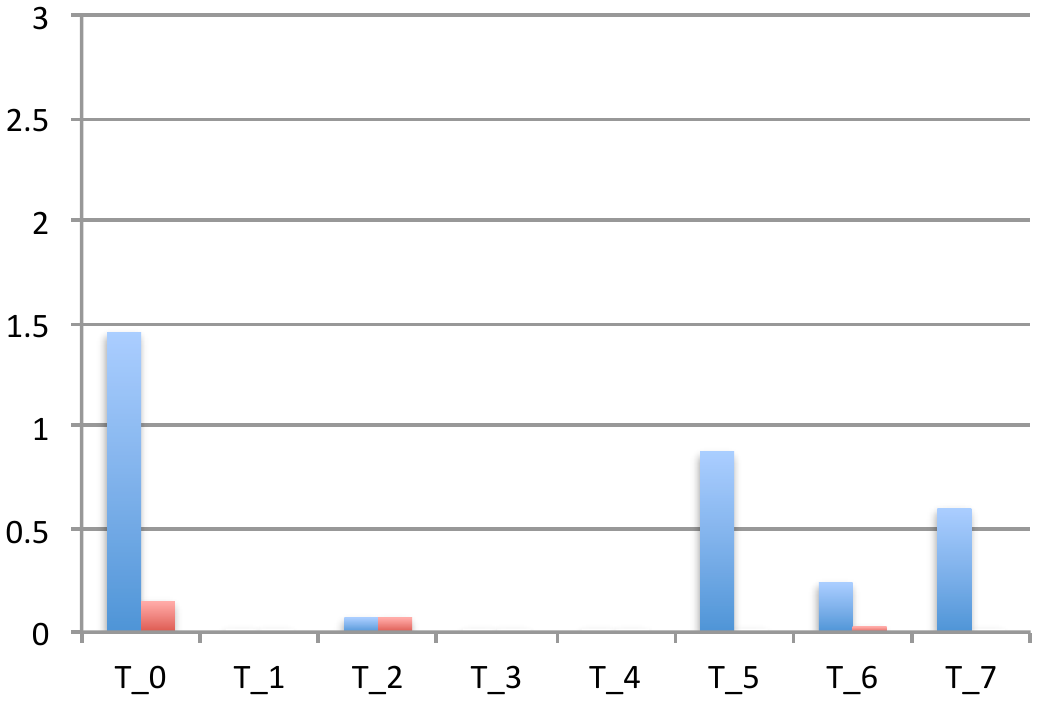}
\\ 
$SO(9)\times SU(2)^2$ lattice 
\\[1ex] 
\includegraphics[width=0.45\textwidth]{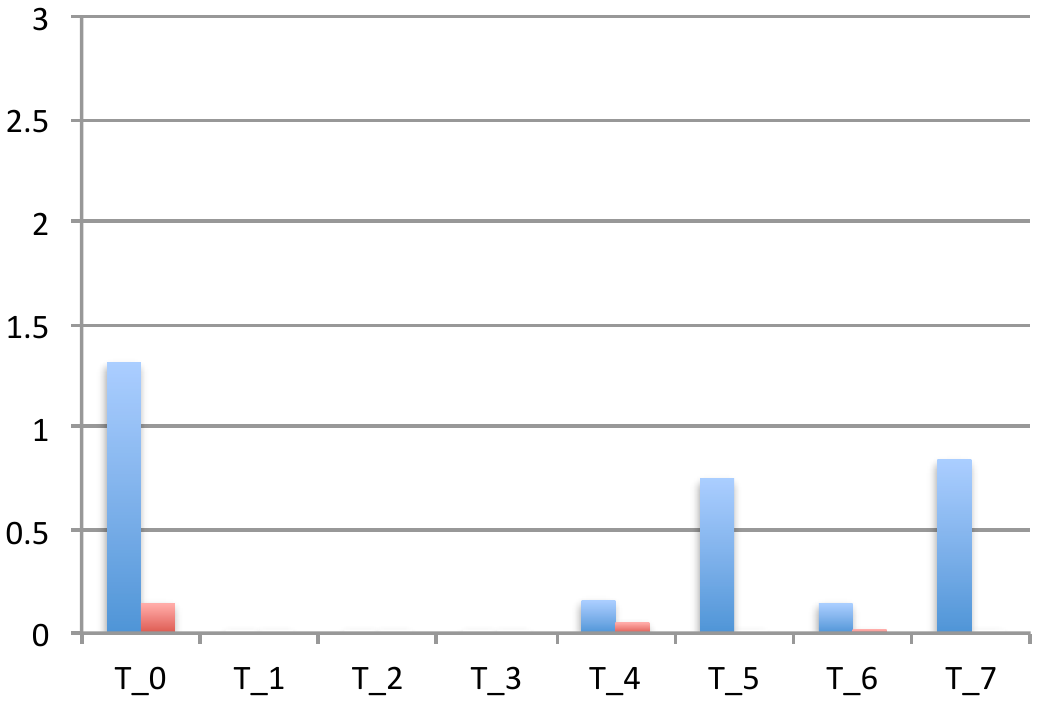}
\\
$SO(10)\times SU(2)$ lattice
\\[1ex] 
\includegraphics[width=0.07\textwidth]{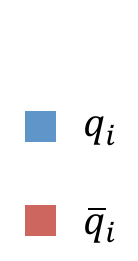}
\vspace{34.5mm}
}
}
\end{figure}
}
%

%
\newcommand{\DistrHiggses}{ 
\begin{figure}[ht!!]
\caption{
These histograms depict the average lepton and Higgs distributions over the different sectors for each orbifold. The three SM leptons and the additional down--type Higgses are collectively denoted by $l_i$ and are plotted in blue whereas the up--type Higgses $h^+_i$ in red.
\label{Higgs_histograms_SUSO}}
\hspace{-1ex} 
\tabu{cc}{
\tabu{c}{
(a) $\mathbb{Z}_\text{8--I}$ models
\\[1ex] 
\includegraphics[width=0.45\textwidth]{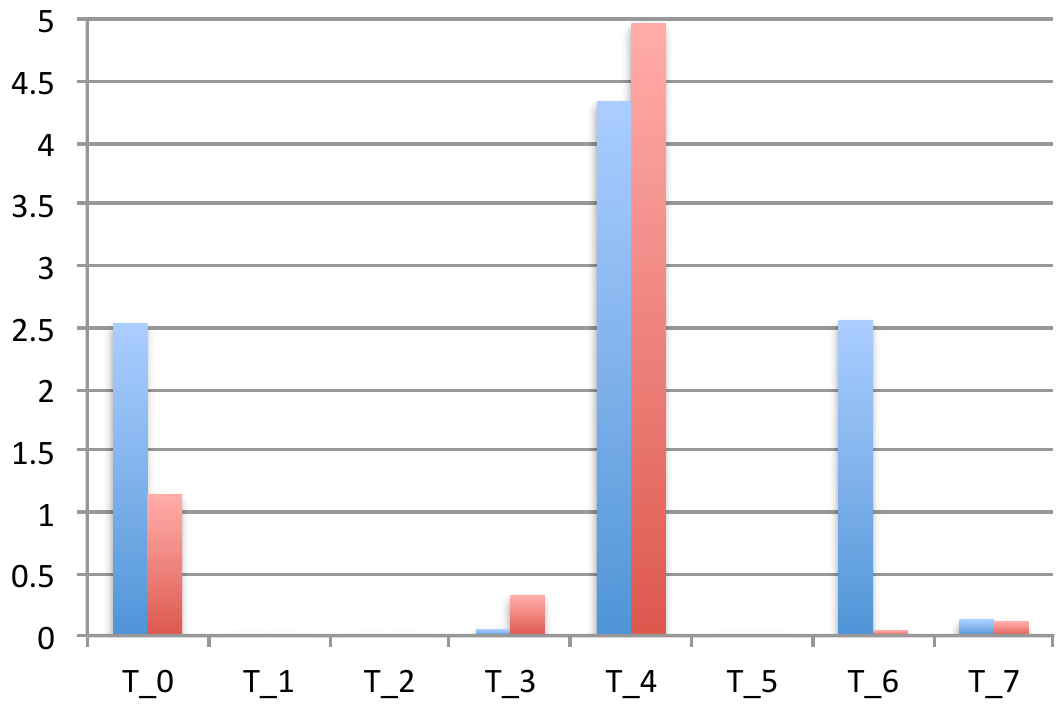}
\\
$SO(9)\times SO(5)$ lattice 
\\[1ex] 
\includegraphics[width=0.45\textwidth]{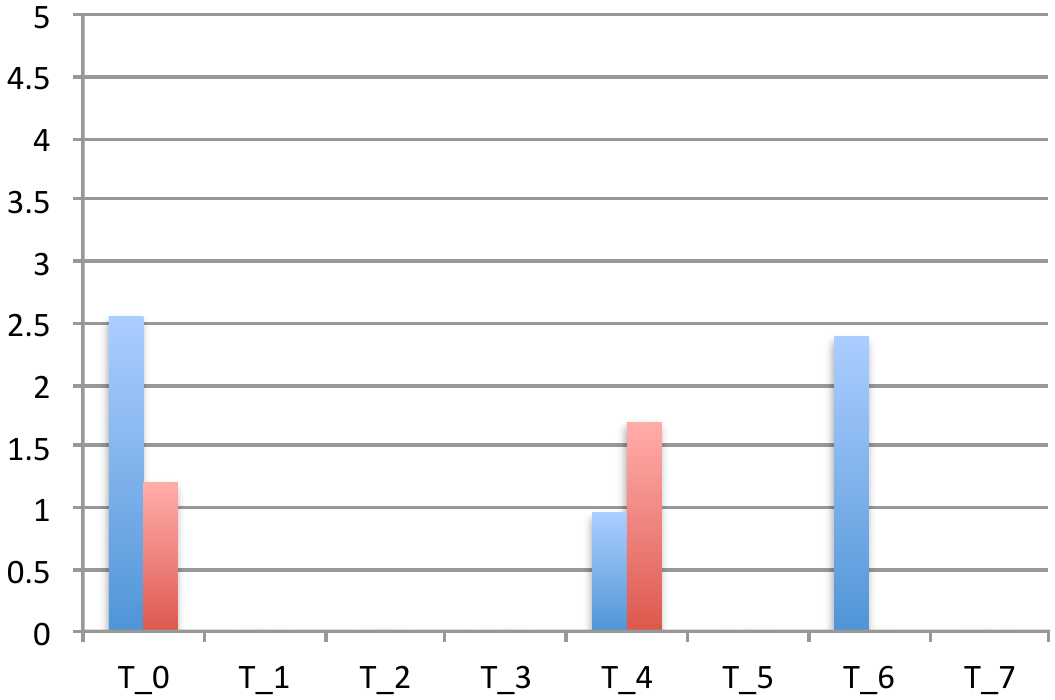}
\\
$SU(4)\times SU(4)$ lattice 
\\[1ex]  
\includegraphics[width=0.45\textwidth]{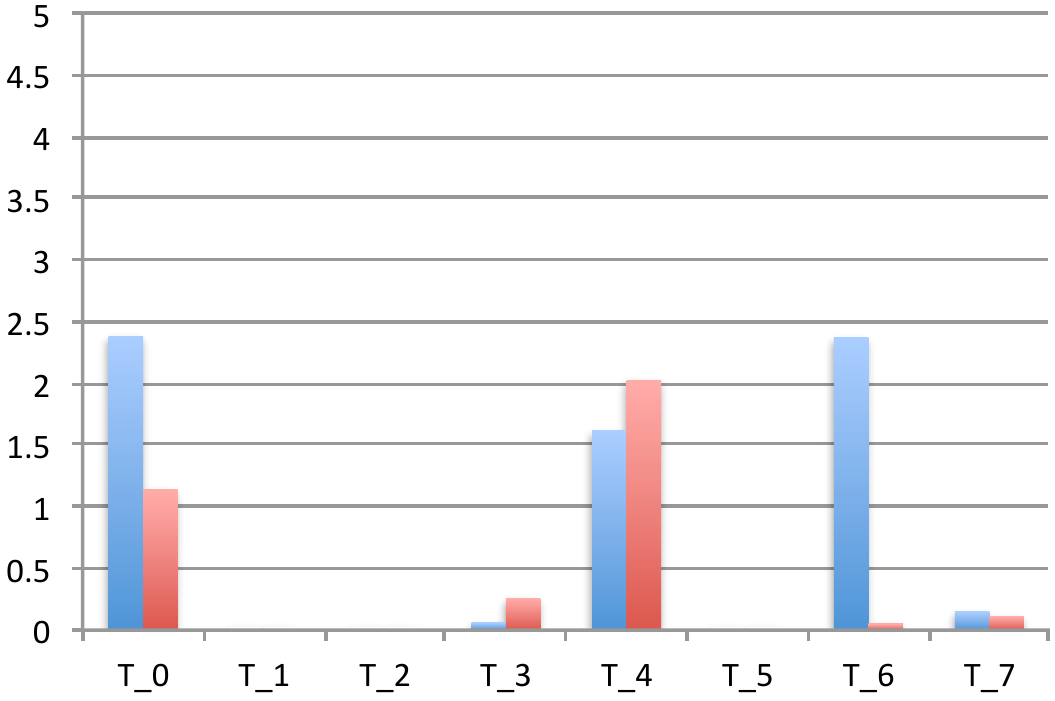}
\\
non--Lie lattice 
}
\tabu{c}{
(b) $\mathbb{Z}_\text{8--II}$ models
\\[1ex] 
\includegraphics[width=0.45\textwidth]{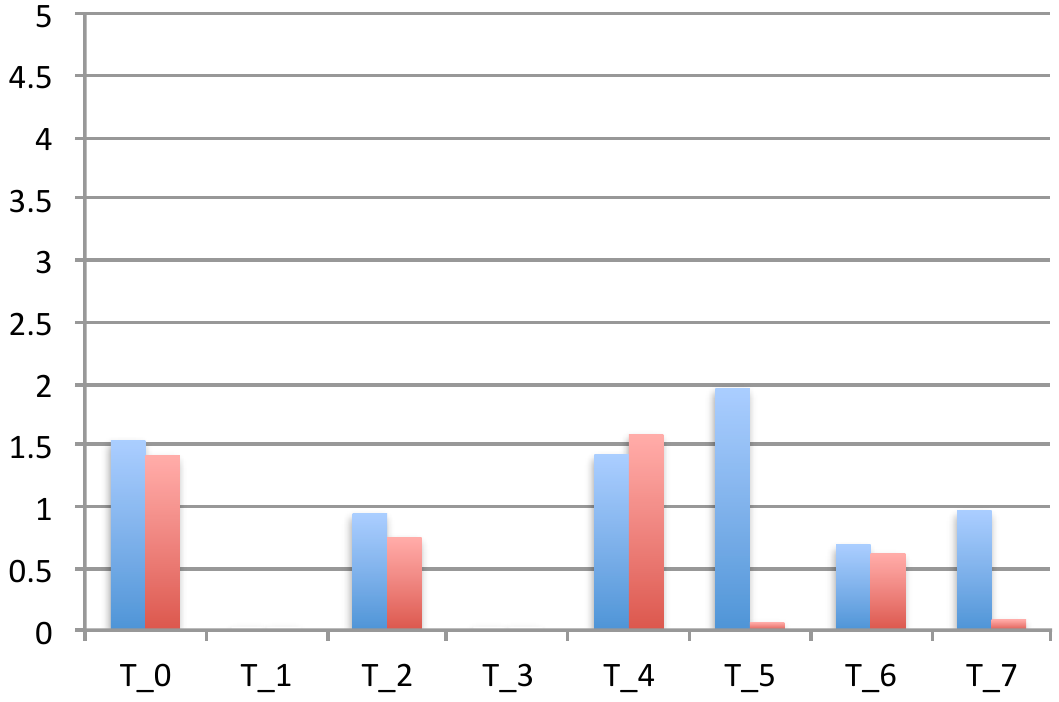}
\\ 
$SO(9)\times SU(2)^2$ lattice
\\[1ex] 
\includegraphics[width=0.45\textwidth]{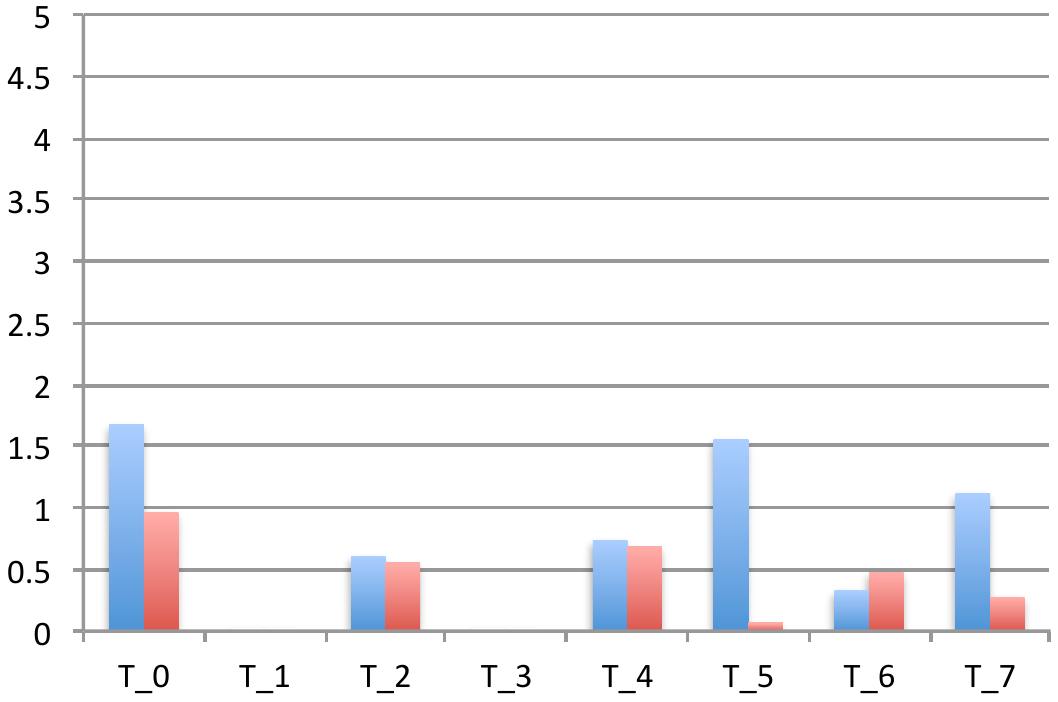}
\\
$SO(10)\times SU(2)$ lattice
\\[1ex]
\includegraphics[width=0.07\textwidth]{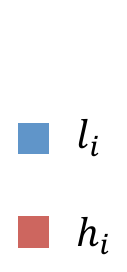}
\vspace{29.0mm}
}
}
\end{figure}
}
%

%
\newcommand{\DistrExotics}{ 
\begin{figure}[ht!!]
\caption{These histograms display the averages of the relative exotic distributions over the various twisted sectors on each orbifold. Here, the states $s_i^+, s_i^-$  and $s_i^0$ are depicted in blue, red and green, respectively. \label{Exotics_histograms_SOSU}}
\hspace{-1ex} 
\tabu{cc}{
\tabu{c}{
(a) $\mathbb{Z}_\text{8--I}$ models
\\[1ex] 
\includegraphics[width=0.45\textwidth]{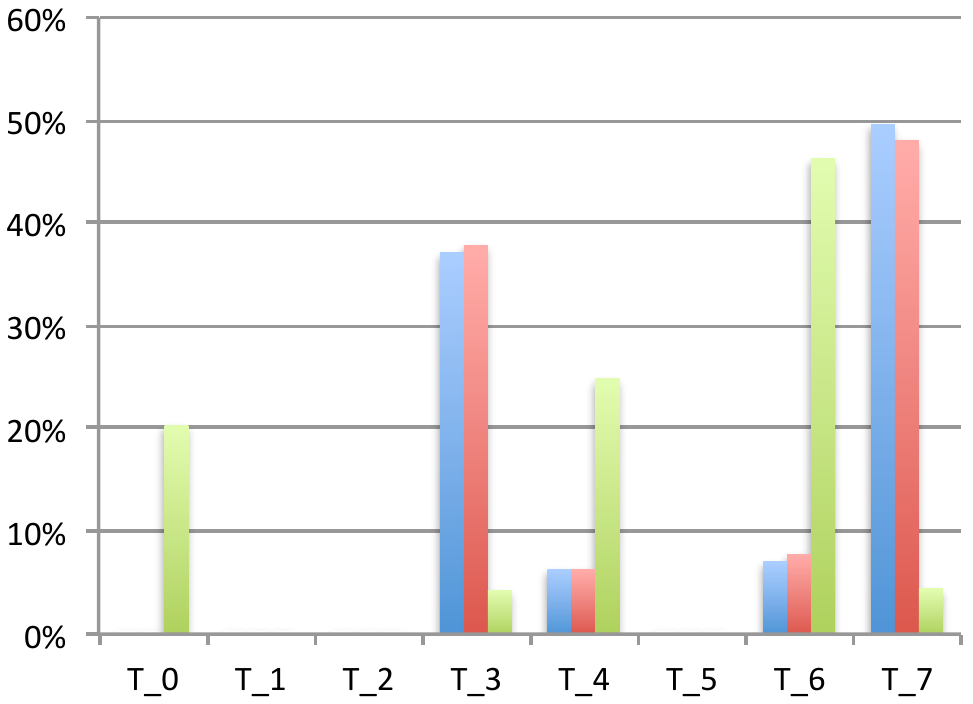}
\\
$SO(9)\times SO(5)$ lattice 
\\[1ex] 
\includegraphics[width=0.45\textwidth]{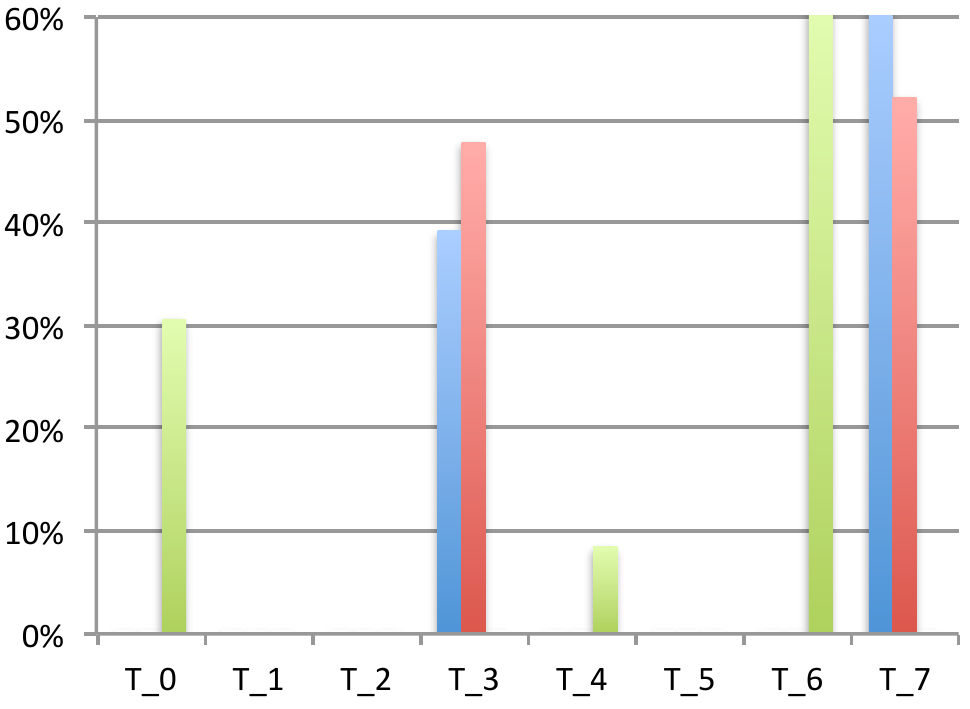}
\\
$SU(4)\times SU(4)$ lattice 
\\[1ex]  
\includegraphics[width=0.45\textwidth]{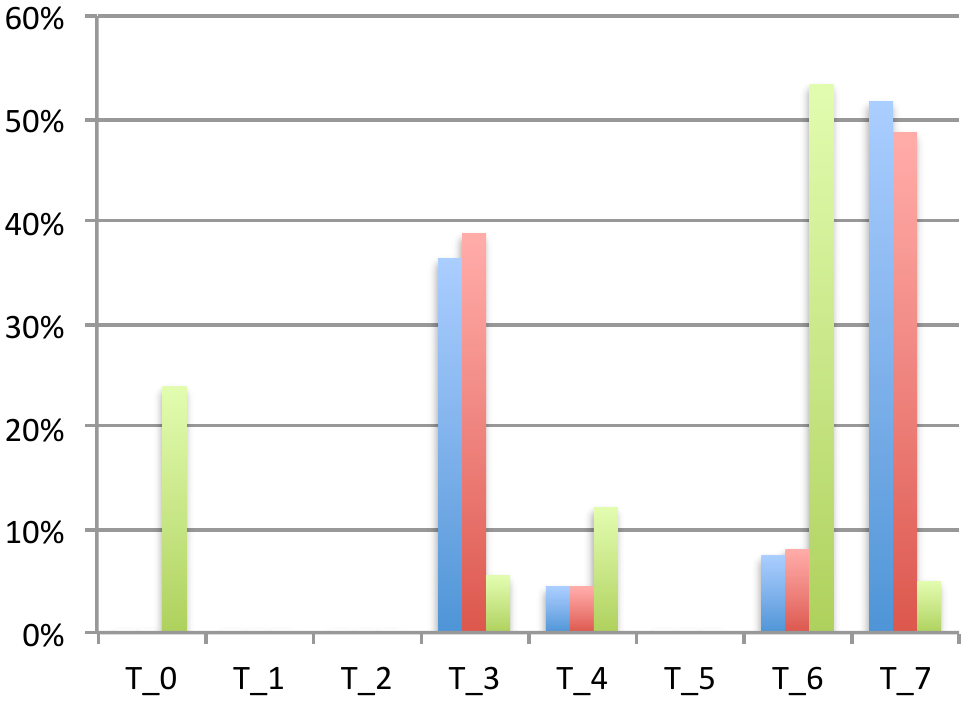}
\\
non--Lie lattice 
}
\tabu{c}{
(b) $\mathbb{Z}_\text{8--II}$ models
\\[1ex] 
\includegraphics[width=0.45\textwidth]{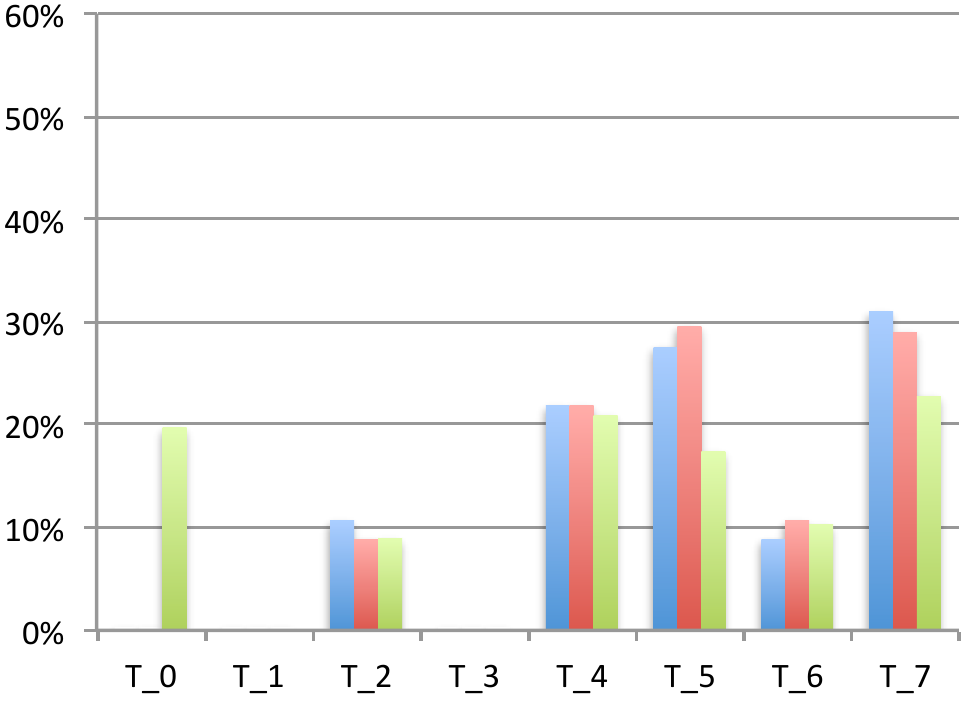}
\\ 
$SO(9)\times SU(2)^2$ lattice
\\[1ex] 
\includegraphics[width=0.45\textwidth]{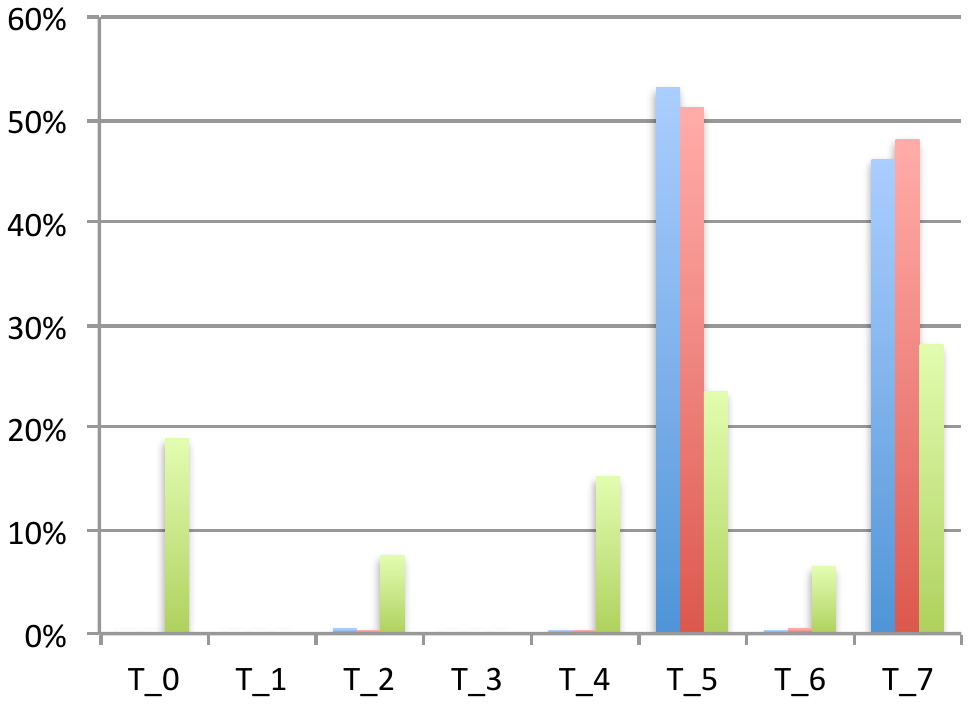}
\\
$SO(10)\times SU(2)$ lattice
\\[1ex]
\includegraphics[width=0.1\textwidth]{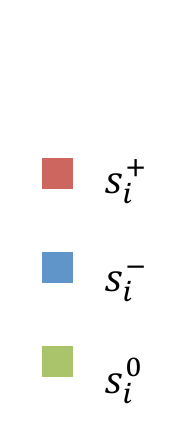}
\vspace{22.0mm}
}
}
\end{figure}
}

\section{Model searches}
\label{sc:models} 

%
In this Section we outline our procedure to uncover MSSM--like models on $\Intr_8$ orbifolds. It is not too difficult to obtain the SM gauge group, $G_{SM}=SU(3) \times SU(2) \times U(1)_Y$, by suitable choices of gauge shifts and discrete Wilson lines (see e.g.~\cite{Ibanez:1987xa}). The real challenge is to obtain the correct matter spectrum containing three SM generations and a single Higgs--doublet pair. Moreover, since such models are often plagued by exotics, i.e.\ states charged under $G_{SM}$ but not part of the MSSM, one has to ensure that these exotic states can be decoupled from the low--energy theory. A necessary condition for this is that they form vector--like pairs w.r.t.\ the SM gauge group. With this in mind we employ the following search criteria for MSSM--like models: 
\enums{
\item The model has an SM gauge group factor $G_{SM}$,
\item with the correct GUT hyper--charge normalization, 
\item has precisely three net generations of quarks and leptons, 
\item contains at least one vector--like Higgs pair, 
\item all SM exotics are vector--like w.r.t.\ the SM gauge group. 
}

%
Defining when two models are considered to be equivalent represents an important issue in MSSM searches. The concept of equivalence is not entirely unambiguous: Different sets of gauge shifts and discrete Wilson lines may lead to string models with identical partition functions. However, it can be very cumbersome to establish this directly. Therefore, rather than considering full partition functions, one often compares the theories on the level of the effective four dimensional low--energy theories only. However, this ignores the possibility that two different string configurations might lead to identical effective low--energy theories. Even with this simplification one has to face the following practical restriction: The zero--mode spectra generically contain a large number of fields which are charged under a variety of $U(1)$ symmetries. Since the bases of two models for the $U(1)$'s will in general not be the same, comparing the spectra on the level of the $U(1)$ charges can be very complicated. 

%
For these reasons we will consider two models to be equivalent if they have the same non--Abelian gauge symmetries and the same spectra w.r.t.\ these non--Abelian symmetries as well as the $U(1)_Y$ hyper--charge assignment. In particular, it might happen that two models defined from different shifts given in Tables~\ref{shift_commulative} or~\ref{shift_commulative_SO} can result in identical spectra, and hence are only counted once.
In addition, we check whether the model allows for a VEV--configuration such that a consistent hyper--charge assignment is possible, while completely ignoring the remaining $U(1)$'s. In favour of having reasonably fast scans, we accept that this simplification might mean that we occasionally classify two models as equivalent even though their $U(1)$ charge assignments would in fact distinguish them.  

\clearpage 
\ZeightSUShifts 
\clearpage 
\ZeightSOShifts
\clearpage 

\subsection{Selection of gauge shift and Wilson lines} 

%
To set up the model searches for $\Intr_8$ orbifolds, we initially tried to employ the models suggested in Ref.~\cite{Kobayashi2} as promising MSSM candidates. However, the combinations of shifts and Wilson lines given in (6)-(8) and (10) of this reference do not satisfy the conditions listed in~\eqref{ModInv}. They rather fulfil the more relaxed modular invariance conditions, e.g.\ discussed in chapter 5 of Ref.~\cite{Choi2006}, which do not ensure consistent orbifold and GSO projections. Because of this the resulting models are plagued by anomalies: We are unable to satisfy the universality conditions for the anomalous $U(1)$, see e.g.\ Ref.~\cite{Casas}; all combinations of shifts and Wilson lines suggested in Ref.~\cite{Kobayashi2} suffer from more than one anomalous $U(1)$. For this reason we avoid using them as a starting point for our model searches.

%
Nevertheless, to build upon some form of $SU(5)$ GUT unification, we either focus on gauge shifts $V$ which break one $E_8$ directly down to $SU(5)$ or on combinations of an $SO(10)$ shift and a Wilson line. The $SU(5)$ gauge symmetry is further broken down to the SM gauge group by the remaining Wilson line(s). (Since the $\Intr_\text{8--I}$ orbifold on the $SU(4)\times SU(4)$ torus lattice only admits a single Wilson line, see \eqref{WLsSU42}, it is quite surprising that MSSM--like models can be constructed at all on this orbifold.) The possible shifts of order $N=8$ acting on a single $E_8$ lattice are taken from Ref.~\cite{shiftZ8}. Not all sixteen dimensional combinations of two such shifts are modular invariant; the allowed $SU(5)$ and $SO(10)$ gauge shifts are tabulated for the two types of $\mathbb{Z}_\text{8}$ orbifolds in Tables~\ref{shift_commulative} and~\ref{shift_commulative_SO}, respectively. All the models tabulated in these Tables are inequivalent and lead to different massless spectra.

\subsection{Computer--aided search}

%
To find models fulfilling the above criteria we wrote a computer code to automatize the construction and classification processes. As input we use the $SU(5)$ and $SO(10)$ shifts given in Tables~\ref{shift_commulative} and~\ref{shift_commulative_SO} and after that we scan over the possible Wilson lines. Only those combinations that fulfill the modular invariance conditions~\eqref{ModInv} compatible with GSO and orbifold projections are kept. 

%
As such scans require the construction of about $10^{11}$ models, many of which are equivalent, filtering techniques have to be implemented. This comes at the price of loosing possible new independent models, therefore one has to proceed with caution here. In our model quests we only relied on elementary symmetries in order to filter out equivalent combinations of Wilson lines. An example of such a symmetry is the permutation of two or more components of a Wilson line that correspond to shift components with the same value. After that preliminary filtering we call certain routines defined in $C$++ classes of the \textit{Orbifolder} (see for a concise description of this package Ref.~\cite{Orbifolder}) to construct the full low--energy models corresponding to these shifts and Wilson lines. Finally, we check whether there are VEV--configurations that support MSSM vacua. 

%
Even though our model scanning code makes extensive use of certain $C$++ classes of  the \textit{Orbifolder} package~\cite{Orbifolder}, we chose not to use this package as a whole: The \textit{Orbifolder} performs a lot of consistency checks which we have already implemented on our input data. Hence, bypassing these checks improved the scanning speed considerably. A second reason why we refrained from only relying on the \textit{Orbifolder} is that, in the current version, the program ignores the hyper--charge when deciding whether two models are equivalent. Therefore, the \textit{Orbifolder} will consider models as equivalent even though they differ in their hyper--charge assignments. However, we have used the front--end of the \textit{Orbifolder} to perform extensive cross checks on our results. 

\clearpage 
\MSSMsummary

\subsection{MSSM--like models}

%
A summary of the models found on the $\mathbb{Z}_8$ orbifolds is provided in Table~\ref{MSSM_averages_SUSO}. 
In this Table (and the Tables and Figures to follow) we present the results for all five inequivalent $\Intr_8$ orbifolds side by side to facilitate comparison. This Table lists all states charged under the SM group $G_{GM}$ and gives their average multiplicities. These average values are computed w.r.t.\ the total number of models for each of the five inequivalent $\Intr_8$ orbifolds  which are listed in the final column of Table~\ref{MSSM_averages_SUSO}. These averages show that on each of these lattices the effective number of generations is three and there is always at least one Higgs pair, as required by our criteria. 

%
In more detail, the MSSM--like models given in Table~\ref{MSSM_averages_SUSO} exhibit the following features: 
The number of exotic vector--like quark--doublet pairs is relatively small. Moreover, the $SU(4)\times SU(4)$ lattice for the $\Intr_\text{8--I}$ orbifold seems to be special in the sense that all the models have exactly three quark--doublets and three right--handed electrons. In fact, for all lattices the number of models with exactly three generations of quark--doublets is very large (about $87\%$) as can be seen from Table~\ref{Quarks_summary_SUSO}. This explains why the average values for the number of quark--doublets in Table~\ref{MSSM_averages_SUSO} is so close to three. In particular, all 49 $SU(4)\times SU(4)$ models and almost all (78 out of 81) non--Lie models have exactly three generations of quark--doublets. Furthermore, we see that the number of additional vector--like $\bar u_i, u_i$--quark pairs and $\bar e_i,e_i$--electron pairs is quite small on all lattices. On the other hand, the number of vector--like pairs of $\bar d_i, d_i$--quarks and the number of Higgs pairs are quite sizable (of the order of three or more) for all inequivalent $\Intr_8$ orbifolds. 
%
%
Interestingly, our set of MSSM--like models contains two models on $SU(4)\times SU(4)$ $\Intr_\text{8--I}$ orbifold that precisely possess three generations of quarks and leptons and one Higgs pair without any additional MSSM--duplicate vector--like pairs.

\clearpage 
\HiddenGauge
\ExactThreeGen
\clearpage

%
However, even these special models contain many exotics. Generally, as can be inferred from Table~\ref{MSSM_averages_SUSO}, the vector--like exotics with different hyper--charges than those of SM states demonstrate following characteristic patterns: The bi--fundamental pairs, $y_i$, $\bar y_i$, are very rare for all five $\Intr_8$ orbifolds, while the triplet pairs, $\phi_i$, $\bar \phi_i$, appear frequently. Table~\ref{MSSM_averages_SUSO} shows that the neutral doublets, $w_i$, singlets, $s^0_i$, and the charged singlets, $s_i^\pm$, can be abundantly found on any of these orbifolds. However, one should realize that in this Table the states are only classified w.r.t.\ their SM charges; whether they are charged under the hidden gauge group is ignored. 

%
This brings us to another interesting point concerning MSSM--like models, namely, the types of gauge groups that arise in the hidden sector. The frequencies, with which various hidden gauge groups appear in the search, are displayed in Table~\ref{hidden_sectors_SUSO} for the five $\Intr_8$ orbifolds under consideration. Given that the hidden groups are big provides an explanation for the large number of $s_i^0, s_i^\pm$ and $w_i$ states: Since some of these states form sizable representations under the hidden group, their multiplicities, ignoring the hidden group, will appear to be large. To appreciate this, one has to realize that $\sim\! 30\%$ of $s^0_i$ in every model is charged under some gauge group factor in the hidden sector. Moreover, large hidden gauge groups could accommodate for supersymmetry breaking via gaugino condensation~\cite{Nilles:1982ik,Ferrara:1982qs}. Hence, also in this respect the $\Intr_8$ models seem appealing.

\subsubsection*{Distribution of states over the different orbifold sectors}

%
Not only the number of states but also their distribution among the orbifold sectors is of interest for phenomenology. For example, a bulk up--type Higgs $h^+$, which can freely move on the orbifold, could help explain the large top--Yukawa coupling observed in the Standard Model. Yukawa couplings are exponentially suppressed for the fields that are located at different fixed points. Therefore, in order to have a natural explanation why the Yukawa coupling of the third generation is of the order of the gauge couplings at the GUT scale while the others are much smaller, it is beneficial when one quark--doublet and a Higgs--pair live in the bulk. That is, they exclusively come from the untwisted sector, while the remaining quark--doublets are localized at fixed points.

%
Table~\ref{Quarks_summary_SUSO} indicates how many models for each of the $\Intr_8$ lattices have one, two or three families in the untwisted sector. Moreover, the histograms in Figure~\ref{Quark_distribution_SUSO} display the distributions of the average number of quark--doublet multiplets over the untwisted and twisted sectors, denoted as T\_0 and T\_1 through T\_7, respectively. The averages are taken for each of the five inequivalent $\Intr_8$ orbifold lattices separately. Moreover, the fact that the net number of quark--doublets is input has been used to fix the normalization: The sum of all $q_i$ (blue) histograms minus the sum of all $\bar q_i$ (red) histograms is always 3. In models with more than three quark multiplets the extra unwanted families can be paired up with their conjugates provided that either the pairing takes place at the same fixed point or some of these states are free to move along some torus directions to reach their localized partners. As can be inferred from Table~\ref{Quarks_summary_SUSO}, a large portion of the $SO(9)\times SU(2)^2$ models have exactly three quark--doublets; most of them have one untwisted and two twisted quark--doublets.

%
We have used the same tactic to plot the Higgs distributions in Figure~\ref{Higgs_histograms_SUSO}.
They clearly show that the $SO(9)\times SO(5)$ orbifold has more vector--like Higgs pairs than any of the other $\Intr_8$ orbifolds. 
Inspecting histograms displayed in Figure~\ref{Quark_distribution_SUSO} and~\ref{Higgs_histograms_SUSO} together could provide an explanation for the top--quark Yukawa hierarchy problem. On $\mathbb{Z}_\text{8--II}$ lattices there are more models potentially leading to a large top mass: 33 models on $SO(9) \times SU(2)^2$ and 8 on $SO(10) \times SU(2)$ exhibit exactly one untwisted SM family without any MSSM vector--like exotics. In addition, these models include one untwisted up--type Higgs $h^+$ so that the top--quark Yukawa coupling can be realized entirely in the bulk.

\clearpage
\DistrQuarks
\clearpage
\DistrHiggses
\clearpage 
\DistrExotics 
\clearpage

%
Finally, let us discuss the distribution of vector--like exotics in our MSSM scans on the $\Intr_8$ orbifolds. Exotic states come exclusively from twisted sectors, except for some of the $s^0_i$ multiplets given in 
Table~\ref{MSSM_averages_SUSO}. At least $\sim\! 20 \%$ of $s^0_i$ states in every model can freely move in the bulk. Among them we could identify then, the uncharged moduli from the untwisted sector. The $s_i^\pm$ states mostly appear in pairs in each twisted sector so that they could immediately pair--up. Despite that, some $s_i^\pm$--pairs reside in the fixed points of different but through anti--twist (same fixed set configuration) related sectors, see the histograms in Figure~\ref{Exotics_histograms_SOSU}. Contrary to the previous histograms, we have plotted the exotics percentage--wise, because the actual number of exotic states vastly varies over the inequivalent models on the compatible lattices.